\documentclass[aoas,preprint]{imsart}
\RequirePackage{amsthm,amsmath,amsfonts,amssymb}
\RequirePackage[authoryear]{natbib}
\RequirePackage[colorlinks,citecolor=blue,urlcolor=blue]{hyperref}
\RequirePackage{graphicx}
\usepackage[utf8]{inputenc}
\usepackage{amsmath}
\usepackage{amsfonts}
\usepackage{amssymb}
\usepackage{mathtools}
\usepackage{enumitem}
\usepackage{tikz}
\usetikzlibrary{fit,positioning}
\graphicspath{ {./STM/} }
\usepackage{url}
\usepackage{subcaption}
\usepackage{multirow}
\usepackage{bbm}
\usepackage{tabulary}
\usepackage{float}
\usepackage{color,soul}
\usepackage{footnote}
\usepackage{natbib}
\usepackage{comment}
\usepackage{algorithm}
\usepackage{algpseudocode}
\makesavenoteenv{tabular}

\startlocaldefs

\theoremstyle{remark}

\endlocaldefs

\begin{document}

\begin{frontmatter}
\title{Regional Topics in British Grocery Retail Transactions}
\runtitle{Regional Topics in British Grocery Retail Transactions}

\begin{aug}
\author[A]{\fnms{Mariflor} \snm{Vega-Carrasco}\ead[label=e1]{mariflor.vega.15@ucl.ac.uk}},
\author[A]{\fnms{Mirco} \snm{Musolesi}\ead[label=e2,mark]{m.musolesi@ucl.ac.uk}}
\author[B]{\fnms{Jason} \snm{O'Sullivan}\ead[label=e3,mark]{jason.osullivan@dunnhumby.uk}}
\author[B]{\fnms{Rosie} \snm{Prior}\ead[label=e4,mark]{rosie.prior@dunnhumby.uk}}
\and
\author[C]{\fnms{Ioanna} \snm{Manolopoulou}\ead[label=e5,mark]{i.manolopoulou@ucl.ac.uk}}
\address[A]{Department of Computer Science, University College London,
\printead{e1,e2}}

\address[B]{Dunnhumby,
\printead{e3,e4}}

\address[C]{Department of Statistical Sciences, University College London,
\printead{e5}}

\end{aug}

\begin{abstract}
Understanding the customer behaviours behind transactional data has high commercial value in the grocery retail industry. Customers generate millions of transactions every day, choosing and buying products to satisfy specific shopping needs. Product availability may vary geographically due to local demand and local supply, thus driving the importance of analysing transactions within their corresponding store and regional context. Topic models provide a powerful tool in the analysis of transactional data, identifying topics that display frequently-bought-together products and summarising transactions as mixtures of topics. We use the Segmented Topic Model (STM) to capture customer behaviours that are nested within stores. STM not only provides topics and transaction summaries but also topical summaries at the store level that can be used to identify regional topics. We summarised the posterior distribution of STM by post-processing multiple posterior samples and selecting semantic modes represented as recurrent topics. We use linear Gaussian process regression to model topic prevalence across British territory while accounting for spatial autocorrelation. We implement our methods on a dataset of transactional data from a major UK grocery retailer and demonstrate that shopping behaviours may vary regionally and nearby stores tend to exhibit similar regional demand. 
\end{abstract}


\end{frontmatter}

\section{Introduction}

In the grocery retail industry, millions of transactions are generated every day by customers that choose and buy products to fulfil one or more needs. Transactions typically contain few products out of thousands of available items, reflecting the unseen customer motivations. For instance, customers go to the grocery retailers to buy foods for breakfast, ingredients to cook a roast dinner or popular products for a barbecue. Identifying customer behaviours provides insights into high-resolution shopping patterns that may help retailers to maximise efficiency while delivering value to all stakeholders. 


Customer motivations may be driven by geographical effects, i.e., showing product combinations that are only relevant at specific stores. For example, a store in Scotland may offer products from local brands and/or products that are part of the local cuisine; these products might not have the same popularity in other further constituent countries in the UK. In response, retailers customise product assortments to include locally supplied products and to fulfil local demand. Thus, we cannot generalise the prevalence of customer behaviours over a large territory without accounting for spatial variation. 

In the UK, spatial analysis has been previously applied to grocery retail data to study store catchment and store performance. For example, \cite{sturley2018evaluating} used an agent-based model to extract key consumer behaviours about shopping frequency, shopping mission, store choice and spending. \cite{davies2019buy} applied a spatial interaction modelling (SIM) technique, to create catchment areas and investigate the spatial variation on competition, sales area, trade intensity, among other factors. With a SIM approach, \cite{newing2015developing} forecasted store patronage and store revenues in two English regions. \cite{waddington2018open} explored spatiotemporal fluctuations of store sales and catchment areas. \cite{berry2016using} examined workplace geographies and census statistics to investigate store trading characteristics in inner London. However, none of the existing literature investigates spatial variations of customer behaviours by modelling product combinations directly.

On the other hand, regional food consumption has been discussed in anthropological and sociological works. For instance, \cite{kuznesof1997regional} found that `regional foods' are perceived as `regional products' or `regional recipes', which are associated with high‐value, speciality, or hand‐crafted products and with dishes that require home preparation and cooking. \cite{groves2005local} defined `regional food' as the food of a particular area of the country, often representing a regional speciality. However, these studies were not carried out using transactional data, but instead employed market research methods such as focus groups and questionnaires. 

This paper identifies regional topics and their geographical distributions exploiting grocery transactions in a Bayesian modelling framework. Regional topics represented as product combinations reflect customer behaviours, which may help retailers launch marketing campaigns, customise store assortments and layout, and may also support the prediction of thematic composition for new stores. Besides, geographical resolution of shopping patterns may aid the investigation of eating habits driven by social and cultural factors that otherwise rely on expensive ad hoc studies. Thus, our overarching goal is to identify combinations of products in high demand in specific areas and to determine their spatial prevalence, characterising customer behaviours from different regions and constituent countries of the UK. To this end, two main ingredients are needed: a model for capturing customer behaviours through transactional data, combined with a model that captures the spatial distribution of these customer behaviours.

Customer behaviours can be modelled through topic modelling (TM) applied to transactional data of product combinations. TM is a scalable statistical framework that was originally introduced to analyse and summarise large collections of text corpora. In retail analytics, TM allows describing transactions and groups of transactions as probabilistic mixtures of topics, which are distributions over a fixed product assortment. Different topics exhibit different combinations of products with high probability, expressing different customer behaviours. Topic models have been applied to model customer behaviour over highly aggregated product assortments  \citep{christidis2010exploring,hruschka2014linking,jacobs2016model,hruschka2016hidden,schroder2017using}, but were only recently applied to transactional data at the full product resolution \citep{vegacarrasco2020modelling,hornsby2019conceptual}.

We capture topics with geographical variability by accounting for the dependency of transactions on store-specific product assortment, i.e., transactions can only contain products that are available at their associated stores. Modelling this dependency implies a hierarchy in which stores are one level above transactions. Without store hierarchy, regionally purchased products would be drowned out by the sheer volume of nationally supplied products, hampering the identification of regional topics. Thus, we apply the segmented topic model (STM) \citep{du2010segmented}, which enables the identification of product combinations within the store context. STM provides topic distributions, transaction-specific topical mixtures, and store-specific topical mixtures. Topic distributions describe products that are frequently bought together with high probabilities, reflecting different customer behaviours. Topical mixtures summarise purchased products according to their topical composition, i.e., a very popular topic in a store-specific topical mixture would show a high probability. 

We characterise the spatial distribution of regional topics using linear Gaussian Process regression (LGPR) \citep{banerjee2014hierarchical,cressie2015statistics,williams2006gaussian} on store-specific topic probabilities, modelling topical prevalence over the UK. LGPR accounts for store meta-data and the geographical proximity between stores, both of which STM does not account for. Specifically, we employ a spatial Gaussian process within a linear model on regional covariates, where spatial dependence is represented by the square exponential covariance function. LGPR allows us to identify and characterise variation in the topic probabilities, which are explained by spatial autocorrelation as well as regional covariates. We demonstrate that the LGPR approach naturally achieves a better out-of-sample predictive behaviour than a linear model by borrowing information from neighbouring stores while affording an interpretable model with quantifiable uncertainty. 

Fully integrating STM and LGPR by assuming a Gaussian Process distribution over store-specific topical mixtures, is feasible but computationally prohibitive. The joint model would be similar to the correlated topic model \citep{blei2006correlated} with two layers of topical mixtures and a covariance matrix defined over geographical distance. We do not pursue this approach as the non-conjugacy of the Gaussian Process poses a challenge for posterior inference. Instead, we feed an LGPR with a topical posterior summary obtained from STM, taking advantage of the closed form Gibbs sampler of STM

Summarising the posterior distribution of STM is needed but it is not an easy task. Topic models are often highly multi-modal, resulting in topics that may not reappear among posterior samples \citep{chuang2015topiccheck, rosen2004author}. Here, we summarise posterior topic distributions by identifying thematic modes following the clustering methodology in \cite{vegacarrasco2020modelling}. This methodology fuses topic distributions from multiple posterior samples to identify recurrent topics and their associated uncertainties. Topics are grouped into clusters, which are represented by their average distribution, named \textit{clustered} topics, and by their cluster size, named \textit{recurrence}. Users evaluate subsets of clustered topics and select a posterior topical summary depending on generalisation and quality metrics. 


This paper is organised as follows: STM and LGPR are introduced in Sections \ref{TM} and \ref{GM}. Regional topics in British grocery retail transactions are presented in Section \ref{topicAnalysis}. Spatial analysis of regional topics is discussed in Section \ref{spatialAnalysis}. Finally, we conclude and summarise our findings in Section \ref{conclusions}.

\section{Topic modelling}\label{TM}

Topic modelling was originally introduced to automatically organise, understand, and summarise large collections of text corpora. Latent Dirichlet Allocation (LDA) \citep{blei2003latent,blei2012probabilistic} is one of the most popular topic modelling techniques, which represents documents as mixtures of topics, and topics as distributions over a fixed vocabulary. The segmented topic model (STM) \citep{du2010segmented} extends LDA to include hierarchical structure within documents, thereby STM represents documents as collections of paragraphs (segments). Both documents and paragraphs are represented as mixtures of topics, where a paragraph-specific topical mixture derives from its document-specific topical mixture. LDA and STM interpret documents as \textit{bags of words}, disregarding word order.

STM has not been applied in retail analytics to the best of our knowledge, but instead has been mainly used in text applications. For instance, STM has been used to match experts with questions \cite{riahi2012finding} and to analyse multi-aspect sentiment in customer reviews \cite{lu2011multi}. We apply STM in the context of grocery retail data, interpreting stores as documents, transactions as segments and topics as distributions over a fixed assortment of products. Transaction-specific topical mixtures derive from the corresponding store-specific topical mixture. Thus, transactions and stores share the space of latent topics. The \textit{bag of words} assumption organically fits the grocery retail domain since products are registered at stores without an inherent order. 

In the standard LDA model, topics display products that are frequently purchased together. If a product is frequently purchased in few stores (and rarely purchased due to unavailability or low preference in the majority of stores), then the product is unlikely to rank highly within a topic. Thus, analysing retail data through LDA might overlook topics that reflect regional or local customer behaviours. In contrast, STM can harness meta information of store hierarchy over transactions. Thereby, product co-occurrence is relative to store context and transactions taking place at the same store are expected to exhibit more similar topical mixtures than transactions from other stores.

\subsection{Segmented topic model}

STM \citep{du2010segmented} consider the following hidden variables: topic distributions, store-specific topical mixtures and transaction-specific topical mixtures. In detail, \(K\) topic distributions, \([\phi_1,....\phi_K]\), are sampled from a Dirichlet distribution governed by hyperparameters \(\boldsymbol{\beta}\); each \(\phi\) is a \(V\)-dimensional vector, and \(V\) is the size of the product assortment. \(D\) store-specific topical mixtures, \(\theta_1,...,\theta_D\), are sampled from a Dirichlet distribution governed by hyperparameters \(\boldsymbol{\alpha}\); each \(\theta\) is a \(K-\)dimensional vector. \(P\) transaction-specific topical mixtures, \(\nu_{1,d},...,\nu_{P,d}\), are sampled from a Poisson-Dirichlet Process distributed with discount parameter \(a\), strength parameter \(b\) and base measure \(\theta_d\); each \(\nu\) is also a \(K\)-dimensional vector.

STM follows a generative process in which each transaction is created by sampling products from topics, which are also sample from a transaction-specific topical mixture. This generative process has two steps. First, a topic assignment \(z_{n,p,d}\) is sampled from a transaction-specific topical mixture \(\nu_{p,d}\). Second, a product \(w_{n,p,d}\) is sampled from the assigned topic distribution \(\phi_{z_{n,p,d}}\), where \(n\) is the \(n^{th}\) item in transaction \(p\) in store \(d\). Mathematically,

\begin{equation}
\label{eq:STMgenerativemodel}
\begin{aligned}
\phi_k  &\sim \textrm{Dirichlet}(\boldsymbol{\beta})\\
\theta_d &\sim \textrm{Dirichlet}(\boldsymbol{\alpha})\\
\nu_{p,d} &\sim \textrm{PDP}(a,b,\theta_d)\\
z_{n,p,d} &\sim \textrm{Multinomial}(\nu_{p,d}) \\
w_{n,p,d} &\sim \textrm{Multinomial}(\phi_{z_{n,p,d}}),
\end{aligned}
\end{equation}

The Poisson-Dirichlet process (PDP) \citep{buntine2010bayesian,ishwaran2001gibbs,pitman1997two} is a generalisation of the Dirichlet Process, also called the Pitman-Yor process. PDP is useful to handle conjugacy between Dirichlet and Multinomial distributions.

\subsubsection{Inference}\label{STMinference}

PDP has a useful representation called the Chinese restaurant process (CRP) \citep{aldous1985exchangeability}. CRP follows an intuitive analogy in which a Chinese restaurant with infinite \textit{table} capacity receive customers who choose to sit around an occupied table or to open a new table; customers sitting around the same table share the same dish. Interpreting the CRP in the retail context, customers are products and dishes are customer behaviours; thus products that fulfil the same customer need are grouped around the same topic. Note that customers and dishes are linked through tables and a dish can be served by multiple tables. Thus, the CRP introduces `table counts', constrained latent variables \(\boldsymbol{t}\), that represent the number of tables serving the same dish. 

Marginalising transaction-specific variables \(\nu\) introduces the constrained latent variables \(\boldsymbol{t}\) and leaves the store-specific variables \(\theta\) in conjugate form. Integrating out topic distributions \(\phi\) and topical mixtures \(\theta, \nu\), the joint conditional distribution of STM is:


\begin{equation}
\label{eq:STMlikelihood}
\begin{aligned}
&p(\mathbf{z},\mathbf{w},\mathbf{t} \mid \boldsymbol{\alpha},\boldsymbol{\beta},a,b) = \\
&\prod_d\frac{\textrm{Beta}_K(\boldsymbol{\alpha}+\sum_p\mathbf{t}_{p,d})}{\textrm{Beta}_K(\boldsymbol{\alpha})}\prod_{p,d}\frac{(b|a)_{\sum_k t_{p,d,k}}}{(b)_{N_{p,d}}}\prod_{p,d,k} S^{N_{k|p,d}}_{t_{p,d,k},a}\prod_k \frac{\textrm{Beta}_V(\boldsymbol{\beta}+\mathbf{N}_k)}{\textrm{Beta}_V(\boldsymbol{\beta})}
\end{aligned}
\end{equation}
where \(t_{p,d,k}\) is the table count for transaction \(p\), store \(d\) and topic \(k\). \(\textrm{Beta}_K(\boldsymbol{\alpha})\) is the \(K-\)dimensional beta function that normalises the Dirichlet distribution; \(\mathbf{t}_{p,d} =[t_{p,d,1},...,t_{p,d,K}]\) is a \(K-\)dimensional vector of  table count; \((x|y)_N\) denotes the Pochhammer symbol; \(N_{p,d}\) size of transaction \(p\) in store \(d\); \(S^N_{M,a}\) is a generalised Stirling number; \(N_{k|p,d}\) number of topic assignments of topic \(k\) in transaction \(p\) in store \(d\). \(\textrm{Beta}_V(\boldsymbol{\beta})\) is \(V\) dimensional beta function that normalises the Dirichlet distribution; \(\mathbf{N}_k= [N_{1|k}, ..., N_{v|k}, ..., N_{V|k}]\) is a \(V-\)dimensional vector of term counts, which is the number of products of type \(v\) assigned to topic \(k\). Detailed definitions of the Pochhammer symbol and generalised Stirling number are explained in \cite{du2010segmented}. 

Due to the intractable computation of marginal probabilities, the posterior distribution of latent variables cannot be computed directly. Thus, inference of STM is solved by a Gibbs sampler. \cite{du2010segmented,buntine2010bayesian} proposed a Gibbs sampler algorithm, which samples topic assignments and table counts. Later, \citep{du2010segmented,buntine2010bayesian} proposed a more effective algorithm that jointly samples topic assignments and `table indicators' for each term. Table indicators are constraint variables that reconstruct table counts through summation. We use this block Gibbs sampler algorithm in our application of STM. See Appendix \ref{BGS} for more inference details.

The block Gibbs sampler algorithm does not explicitly sampled topics \(\boldsymbol{\phi}\), stores-specific topical mixtures \(\boldsymbol{\theta}\) or transaction-specific topical mixtures \(\boldsymbol{\nu}\). Instead, hidden variables are approximated using a posterior sample \(s\) of topic assignments and table counts. Then, hidden variables are approximated by their conditional posterior means:

\begin{equation}
\label{eq:STMestimates1}
\begin{aligned}
\widehat{\theta}^s_{d,k} &= E(\theta^s_{d,k} \mid \textbf{t}^s, \boldsymbol{\alpha})= \frac{\alpha_k+\sum_p t^s_{p,d,k}}{ \alpha + \sum_{p,k} t^s_{p,d,k}},\\
\end{aligned}
\end{equation}
\begin{equation}
\label{eq:STMestimates2}
\begin{aligned}
\widehat{\nu}^s_{p,d,k} &= E(\nu^s_{p,d,k} \mid \textbf{z}^s,\textbf{t}^s, a,b)= \frac{N^s_{p,d,k}-a\times t^s_{p,d,k}}{b+N^s_{p,d}}+ \theta_{d,k}\frac{\sum_k t^s_{p,d,k}\times a+b}{b+N^s_{p,d}},
\end{aligned}
\end{equation}
\begin{equation}
\label{eq:STMestimates3}
\begin{aligned}
\widehat{\phi}^s_{k,v} &= E(\phi^s_{k,v} \mid \textbf{z}^s,\boldsymbol{\beta})=\frac{\beta_v+N^s_{k,v}}{\beta +N^s_{k}},
\end{aligned}
\end{equation}
where \(\alpha = \sum_k^K \alpha_k\) and where \(\beta = \sum_v^V \beta_v\).

\subsection{Summarising topic distributions}

Summarising the posterior distribution of a topic model is challenging because the posterior distribution is often highly multi-modal; resulting in posterior samples that capture different semantic modes. Thus, component-wise posterior averaging may merge topic distributions that respond to different semantic concepts. In addition, topics that capture different semantic modes may appear and disappear across posterior samples of MCMC chains \citep{chuang2015topiccheck, rosen2004author}.

In response, we follow the methodology described in \citep{vegacarrasco2020modelling} to construct a summary of topical modes using multiple posterior samples from various MCMC chains. The methodology clusters topics using a bottom-up hierarchical clustering method. At each step, the algorithm finds the pair of clusters with the lowest cosine distance and merge clusters if their topic distributions come from different samples. The algorithm keeps merging clusters up to a cosine distance threshold. Each resulting cluster is represented by the average topic distribution, named clustered topic, and by the number of topics gathered in the same cluster, named cluster size. Cluster size is a measure of recurrence and denotes (un)certainty, i.e., a topic that has occurred in every posterior sample is highly recurrent, showing no uncertainty. 


\subsection{Evaluation of clustered topics}

Depending on the cosine distance threshold, the clustering algorithm produces a set of clustered topics, which can be selected according to their recurrence. Thus, multiple subsets of clustered topics can be formed by varying cosine distance threshold and recurrence (setting a minimum cluster size).

We evaluate each subset of clustered topics on 4 aspects: generalisation or predictive power of a subset of topics, coherence of individual topics, the distinctiveness of a topic with respect to the other topics in the same posterior sample, and credibility of a topic with respect to the topics from other posterior samples.

Topic coherence, distinctiveness and credibility are measured as described in  \citep{vegacarrasco2020modelling}. Model generalisation, however, is measured by the perplexity of unseen transactions given topics, store-specific topical mixtures and PDP parameters:
\begin{equation}
\label{eq:Perplexity}
\textrm{Perplexity}=-\frac{\log P(\mathbf{w}^\prime_d \mid \Phi,\theta_d, a,b)}{N^\prime},
\end{equation}
where \(\mathbf{w}^\prime_d\) is a set of products in a held-out transaction at store \(d\), \(N^\prime\) is the number of products in \(\mathbf{w}^\prime_d\), \(\Phi=[\phi_1, \phi_2,\ldots,\phi_K]\) the set of inferred topics, \(\theta_d\) is the store-specific topical mixtures associated to store \(d\), \(a\) and \(b\) are the PDP parameters. 

We aim to select a subset of clustered topics that shows low perplexity, gathering topics that are coherent, distinctive and credible (low uncertainty).

\section{Linear Gaussian process regression}\label{GM}

According to Tobler's first law of geography \citep{tobler1970computer}: \textit{`everything is related to everything else, but near things are more related than distant things'}. Thus, we expect that nearby stores show similar shopping patterns and that some specific patterns may be limited to particular geographical areas. STM does not take into account store location or proximity between stores. Although a topic model that simultaneously accommodate store hierarchy over transactions and store location would be mathematically possible, it would be computationally prohibitive at the level of resolution of interest. Instead, we use the summarised posterior distributions of topics obtained from STM and take a spatial modelling approach to capture their geographical structure and regional behaviour. 

\subsection{Model}

A linear regression with a spatial process is defined as:

\begin{equation}
\label{dataModel}
\textbf{Y} = \textbf{X}\boldsymbol{\beta} + \boldsymbol{\eta}+ \boldsymbol{\varepsilon},
\end{equation}
where \(\textbf{Y}\) is the dependent variable, \(\textbf{X}\) is the matrix of \(p\) covariates associated with locations \(\textbf{s}_1, ...,\textbf{s}_n\), \(\boldsymbol{\beta}\) is a \(p\)-dimensional fixed effect, \(\boldsymbol{\eta}\) is a spatial process, which captures spatial residual, and \(\boldsymbol{\epsilon}\) is an independent process, which models pure error, also known as the \textit{nugget} effect. 

The spatial process \(\eta(\textbf{s}_1),...,\eta(\textbf{s}_n)\) is distributed as a zero-mean Gaussian process \(GP(0,C_{\boldsymbol{\eta}})\) with positive definitive covariance matrix \(C_{\boldsymbol{\eta}}\). Residuals \(\varepsilon(\textbf{s}_1),...,\varepsilon(\textbf{s}_n)\) are assumed \(iid\) with \(\epsilon(\textbf{s}_i) \sim N(0,\sigma^2)\). Thus, observations are distributed as:

\begin{equation}
\label{eq:datamodel}
\textbf{Y} \sim N(\textbf{X}\boldsymbol{\beta}, \Sigma),
\end{equation}
where \(\Sigma = C_{\boldsymbol{\eta}}+ \sigma^2I\).

Here, we use the positive definitive square exponential covariance function,

\begin{equation}
\label{eq:COV}
C_{\boldsymbol{\eta}}(\textbf{s}_i,\textbf{s}_j|\alpha,\rho) = \alpha^2 \exp\Big(-\frac{\textrm{dist}(\textbf{s}_i,\textbf{s}_j)^2}{2\rho^2}\Big),
\end{equation}
where parameters \(\alpha\) and \(\rho\) control the amplitude and length-scale of the spatial dependence, respectively. \(\textrm{dist}(\textbf{s}_i,\textbf{s}_j)\) is a measure of distance between locations. 
\subsection{Methods}

Linear Gaussian process regression specified in equation \ref{dataModel} is fitted using Stan \citep{carpenter2017stan}. Stan is a state-of-the-art platform for statistical modelling and high-performance statistical computation. Stan facilitates Bayesian inference by gradient-based sampling techniques such as Hamiltonian Monte Carlo methods \citep{betancourt2017conceptual} and variational inference \citep{blei2017variational}. In our study, the inference is computed by the default Stan algorithm No-U-Turn Sampler (NUTS) \citep{hoffman2014no}. NUTS is an extension of the Hamiltonian Monte Carlo (HMC) algorithm that effectively explores the parameter space by avoiding retaking previously sampling paths in a U-turn style.  

\subsection{Predictions}

Predicted topic probabilities \(\textbf{Y}^\star=[Y^\star(\textbf{s}_1),...,Y^\star(\textbf{s}_n)]\) at new locations \(\textbf{s}^\star_1, ...,\textbf{s}^\star_n\) are distributed as:
\begin{equation}
\label{eq:Ypred}
\textbf{Y}^\star |\textbf{Y},\boldsymbol{\beta},\Theta, \textbf{X}^\star,\textbf{X}  \sim N(
\textbf{X}^\star\boldsymbol{\beta} + \Sigma_{21}\Sigma_{11}^{-1}(\textbf{Y}-\textbf{X}\boldsymbol{\beta}), \Sigma_{22}-\Sigma_{21}\Sigma_{11}^{-1}\Sigma_{12}),
\end{equation}
where \(\textbf{X}^\star\) is the matrix of \(p\) covariates at the new locations. Here, \(\Sigma_{11}\) is the covariance matrix of \(\textbf{s}_1, ...,\textbf{s}_n\) locations, \(\Sigma_{12}=\Sigma_{21}\) the covariance matrix between \(\textbf{s}_1, ...,\textbf{s}_n\) and \(\textbf{s}^\star_1, ...,\textbf{s}^\star_n\), and \(\Sigma_{22}\), covariance matrix of \(\textbf{s}^\star_1, ...,\textbf{s}^\star_n\).


Note that expected topic probabilities \(E(\textbf{Y}^\star)\) are computed by two quantities. The first quantity is obtained by multiplying the covariate matrix by the fixed effects as in multiple linear regression. The second quantity pulls the expected value at a new store towards the values of the nearby stores if spatial dependence is significant. 

\section{Identifying regional grocery topics}\label{topicAnalysis}

We analyse grocery transactions from a major retailer in the UK. Transactions are sampled randomly, covering 100 nationwide superstores between September 2017 and August 2018. Transactions with less than 3 products are filtered out. The training data set contains 36,000 transactions and a total of 392,840 products and the test data set contains 3,600 transactions and a total of 38,621 products. Transactions contain 10 products on average. The product assortment contains 10,000 products, which are the most monthly frequent, ensuring the selection of seasonal and non-seasonal products. We count unique products in transactions, disregarding the quantities of repetitive products. For instance, 5 loose bananas count as 1 product (loose banana). We do not use an equivalent of stop words list (highly frequent terms), as we consider that every product or combination of them tell different customer needs. We disregard transactions with fewer than 3 products assuming that smaller transactions do not have enough products to exhibit a regional topic. No personal customer data were used for this research.

\subsection{STM posterior summary}\label{HCSTM}

We explore STM with 100 topics to capture as many topics as possible without making inference too computationally prohibited. As shown in \citep{vegacarrasco2020modelling}, a topic model with 100 topics identifies a variety of customer behaviours in the domain of our application. Exploring STM with a smaller or larger number of topic is out the scope of this paper.

We use symmetric priors with hyperparameters \(\alpha_k=1000/K\) and \(\beta_v=0.01\), and PDP hyperparameters \(b=3.0\) and \(a=0.5\). We run four MCMC chains for 100,000 iterations with a burn-in of 80,000 iterations, samples were recorded every 5,000 iterations, obtaining 20 thinned posterior samples (five samples for each chain). MCMC trace plots are presented in Appendix \ref{convergence_STM}, where the convergence is satisfactory. 

Posterior topic distributions are summarised by clustering a bag of 2,000 topics obtained from the aforementioned 20 posterior samples. As shown in Appendix \ref{Posterior_STM}, we observe that the subset formed with a minimum cluster size 10 (which represent 50\% of the samples) and a cosine distance threshold \(\geq 0.35\) show greater coherence, credibility and generalization, concurring with \cite{vegacarrasco2020modelling}. Based on these results, we choose this subset which contains 104 clustered topics.

Store-specific topical mixtures are then obtained by training STM with the identified 104 clustered topic distributions. The inference process goes as described in Section \ref{STMinference}, but only Equations \ref{eq:STMestimates1} and \ref{eq:STMestimates2} are updated. A MCMC chain runs with a burn-in period of 1,000 iterations, recording posterior samples with a thin of 500 iterations. The MCMC trace plot in Appendix \ref{convergence_HCSTM} shows satisfactory convergence. We collect 30 posterior samples which are then averaged to estimate store-specific topical mixtures for 500 stores across the UK.

\subsection{Interpreting topic distributions}

We interpret six out of the 104 clustered topics as they capture a clear, interpretable regional pattern. The remaining topics show ubiquitous distributions over the UK. We interpret topics by analysing the product descriptions of the 15 products with the largest probabilities. Topics are manually named after the regional pattern or customer preference reflected on the product descriptions. Note that the illustrated topics appeared consistently across the 20 posterior samples (size = 20), indicating low posterior uncertainty. 

\begin{figure}[H]
        \centering
        \begin{subfigure}[b]{0.32\textwidth}
            \centering 
            \caption{Scottish}
            \includegraphics[width=1\textwidth]{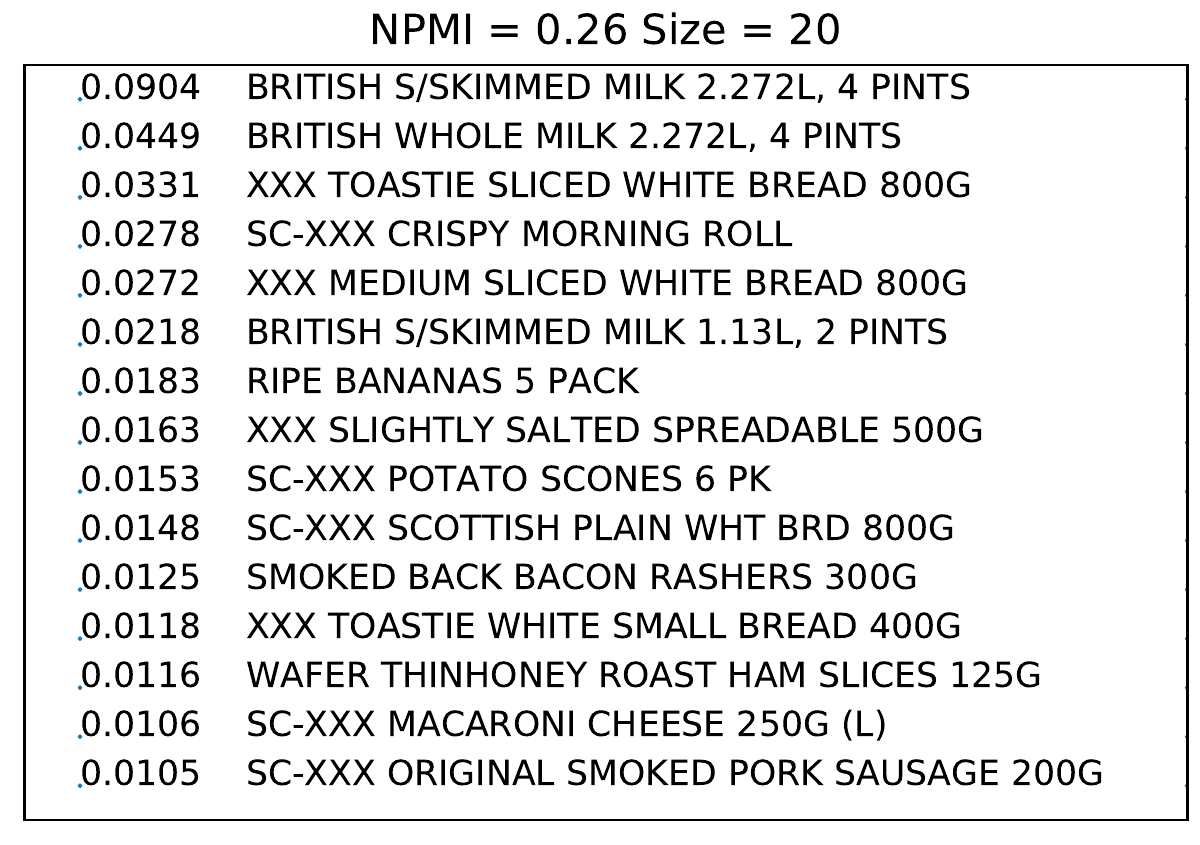}               
            \label{t_2}
        \end{subfigure} 
        \begin{subfigure}[b]{0.32\textwidth}
            \centering 
            \caption{Northern Irish}
            \includegraphics[width=1\textwidth]{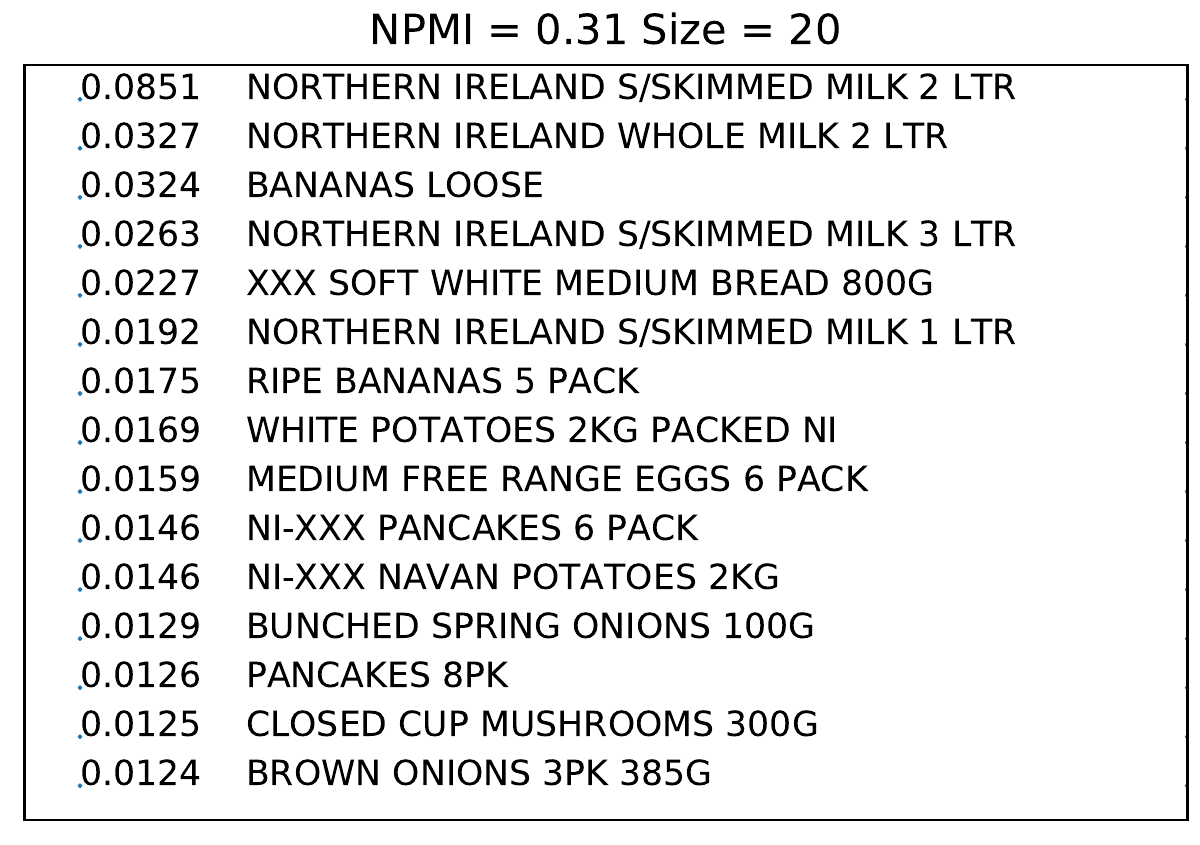}               
            \label{t_1}
        \end{subfigure}    
         \begin{subfigure}[b]{0.32\textwidth}
            \centering 
            \caption{Welsh}
            \includegraphics[width=1\textwidth]{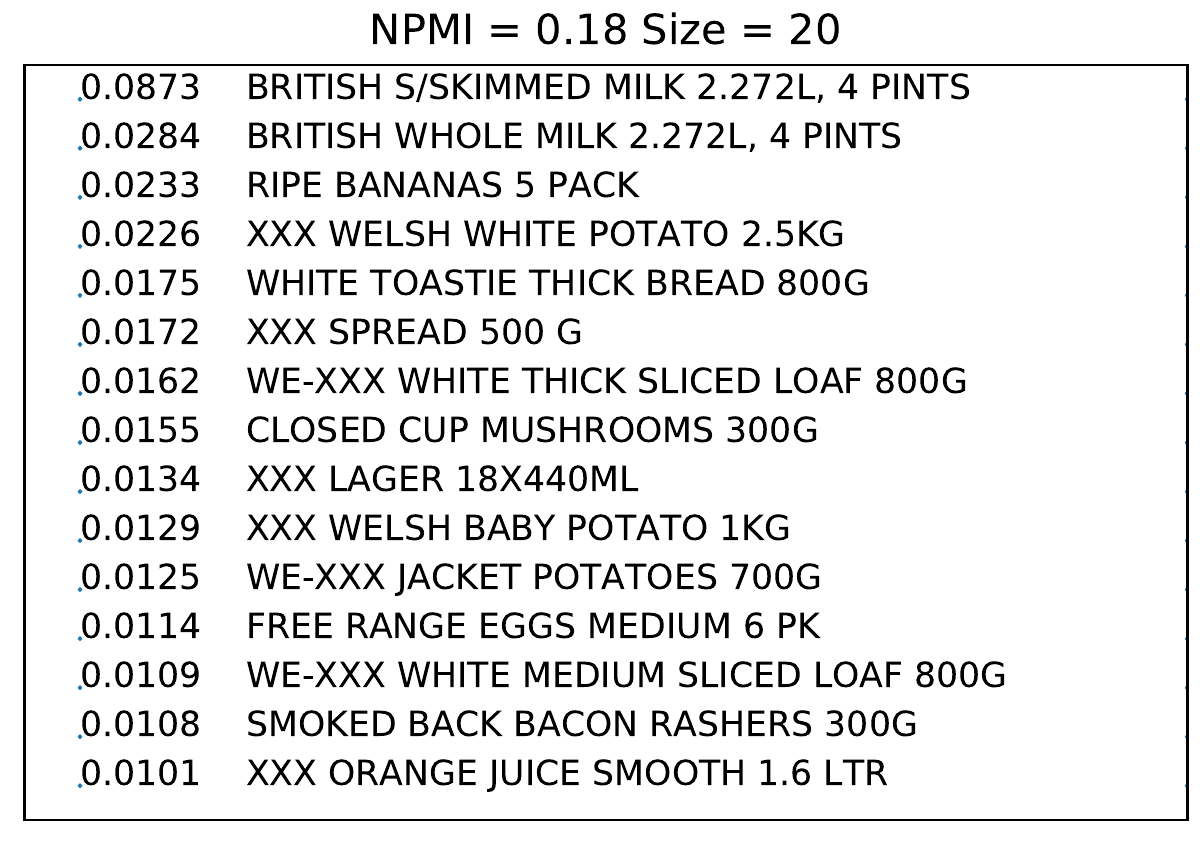}               
            \label{t_3}
             \end{subfigure}          
           \\
        \begin{subfigure}[b]{0.32\textwidth}
            \centering 
            \caption{Mixed basket I}
            \includegraphics[width=1\textwidth]{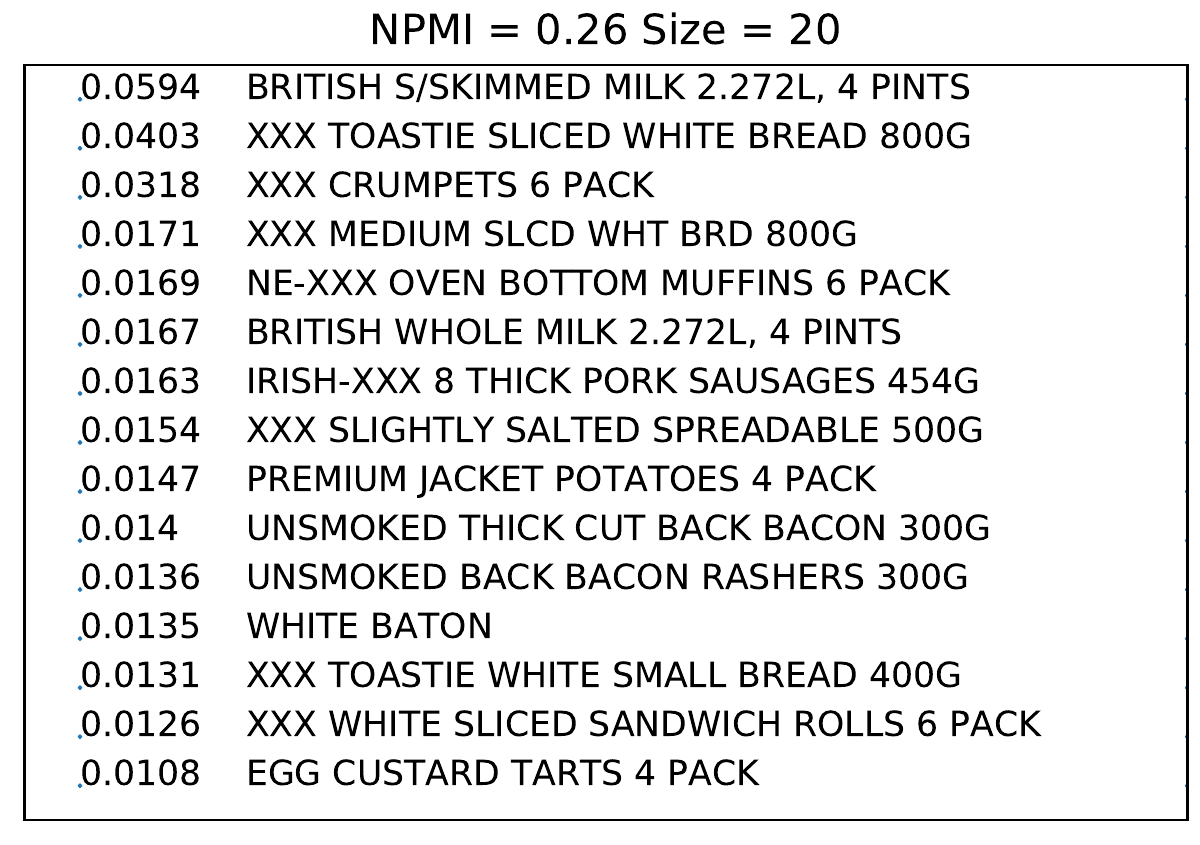}               
            \label{t_4}
        \end{subfigure}
        \begin{subfigure}[b]{0.32\textwidth}
            \centering 
            \caption{Mixed basket II}
            \includegraphics[width=1\textwidth]{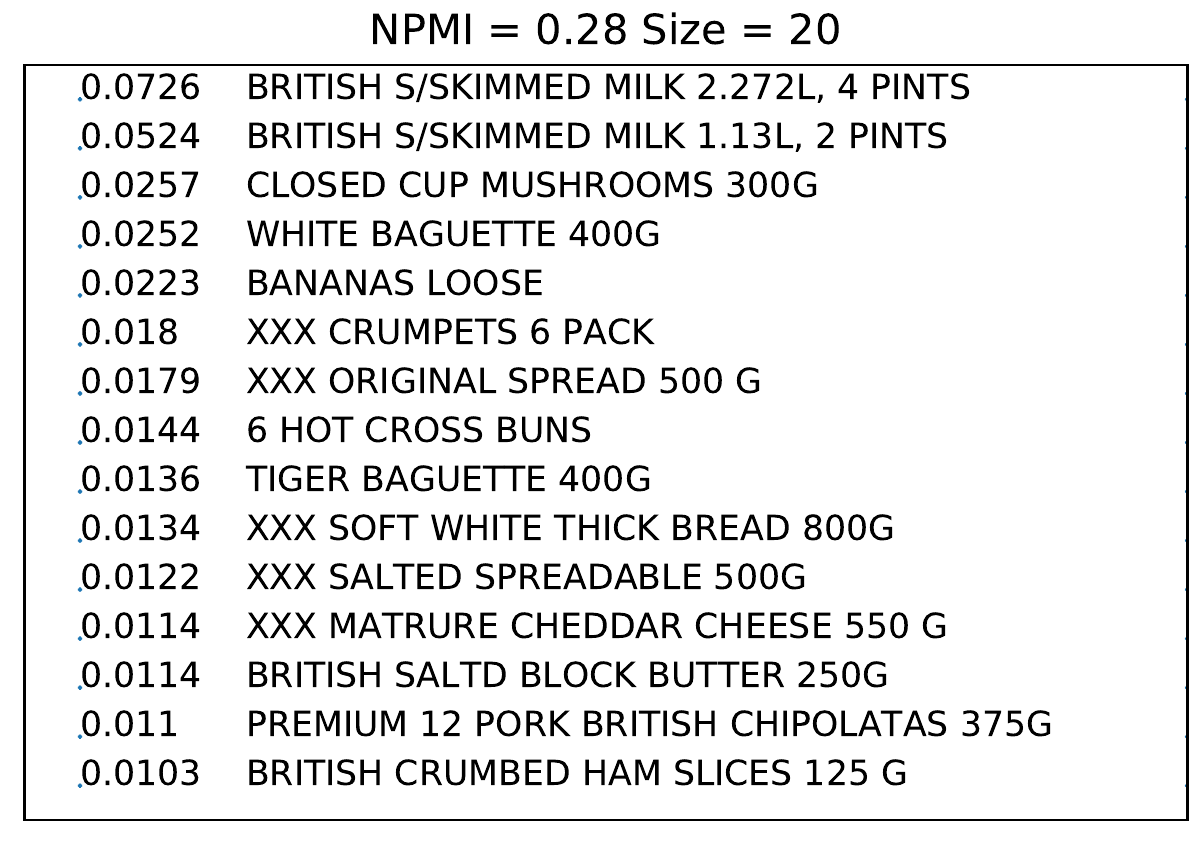}               
            \label{t_5}
        \end{subfigure}
         \begin{subfigure}[b]{0.32\textwidth}
            \centering 
            \caption{Organic}
            \includegraphics[width=1\textwidth]{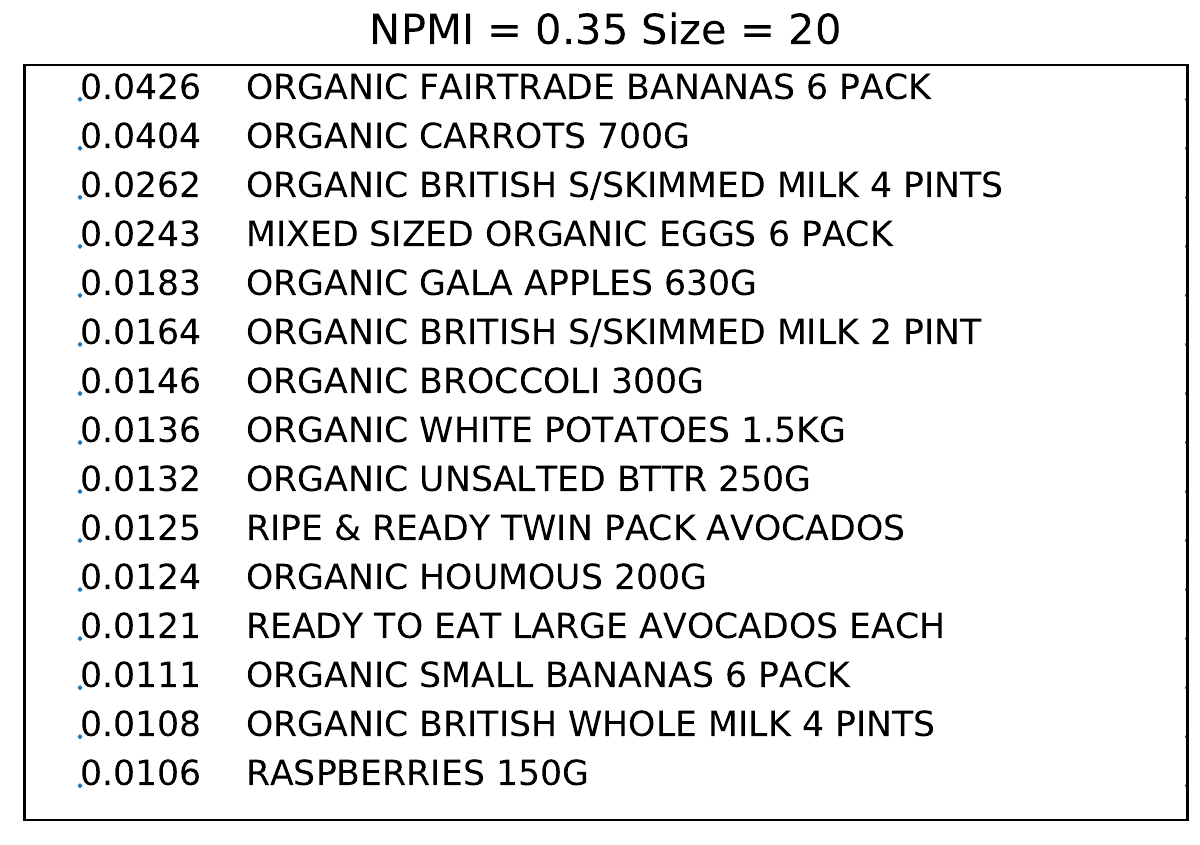}               
            \label{t_6}
        \end{subfigure}      
        \caption{Most probable products in grocery regional topics. Each topic is interpreted using the 15 products with the largest probabilities. Probabilities and products are sorted in descending order. General brand names have been replaced by XXX. Local brands in North Ireland, Scotland, Wales and North of England have been replaced by NI-XXX, SC-XXX, WE-XXX, NE-XXX. NPMI and size are measures of topic coherence and recurrence.}  
        \label{topics_uk}
        \end{figure}

Product descriptions in Figures \ref{t_2}, \ref{t_1} and \ref{t_3} suggest foods supplied locally and local brands associated to Scotland, Northern Ireland and Wales. For instance, the Scottish topic includes `Scottish-branded skinless sausages' and `Scottish-branded potato scones', the Northern Irish topic shows the `North Ireland semi-skimmed milk', `white potatoes packed in North Ireland', and the Welsh topic contains `Welsh jacket potatoes' and `Welsh-branded bread'. Hence, we name the Scottish topic, Northern Irish topic and Welsh topic after the nationality that their product descriptions suggest. 

Figures \ref{t_4} and \ref{t_5} show a variety of products such as types of milk, types of bread, fruits and vegetables, etc. Close inspection of product descriptions such as `oven bottom muffin', `fruit teacake', and `potato and meat pie' in Figure \ref{t_4}, and `pork pies' and `scotch eggs' in Figure \ref{t_5} may reveal a regional topic when regional expertise is available. Since these product descriptions do not provide interpretations that can be directly associated with specific regions, we momentarily name these topics `Mixed basket I' and `Mixed basket II'. Figures \ref{t_6} shows `organic' quality foods, indicating a specific customer preference, however, the topic does not suggest any specific regional pattern. 

Interpreting topic descriptions is not sufficient to identify geographically driven shopping motivations, reinforcing the need for exploring store-specific topical mixtures.

 \begin{figure}
        \centering
        \begin{subfigure}[b]{0.32\textwidth}
            \centering 
            \caption{Scottish}           
            \includegraphics[ width=0.95\textwidth]{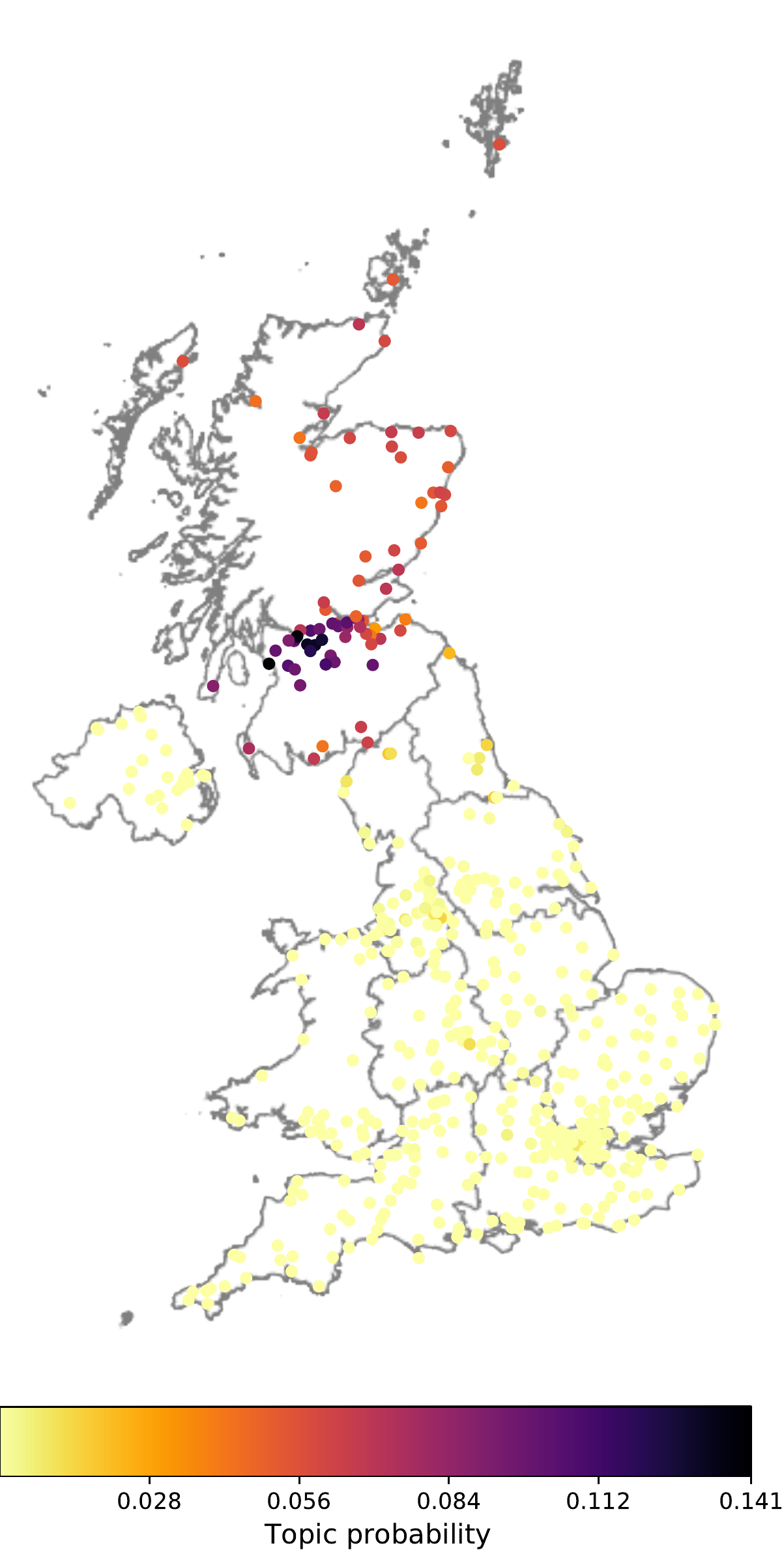}               
            \label{m_1}
        \end{subfigure}
        \begin{subfigure}[b]{0.32\textwidth}
            \centering 
             \caption{Northern Irish}
            \includegraphics[width=0.95\textwidth]{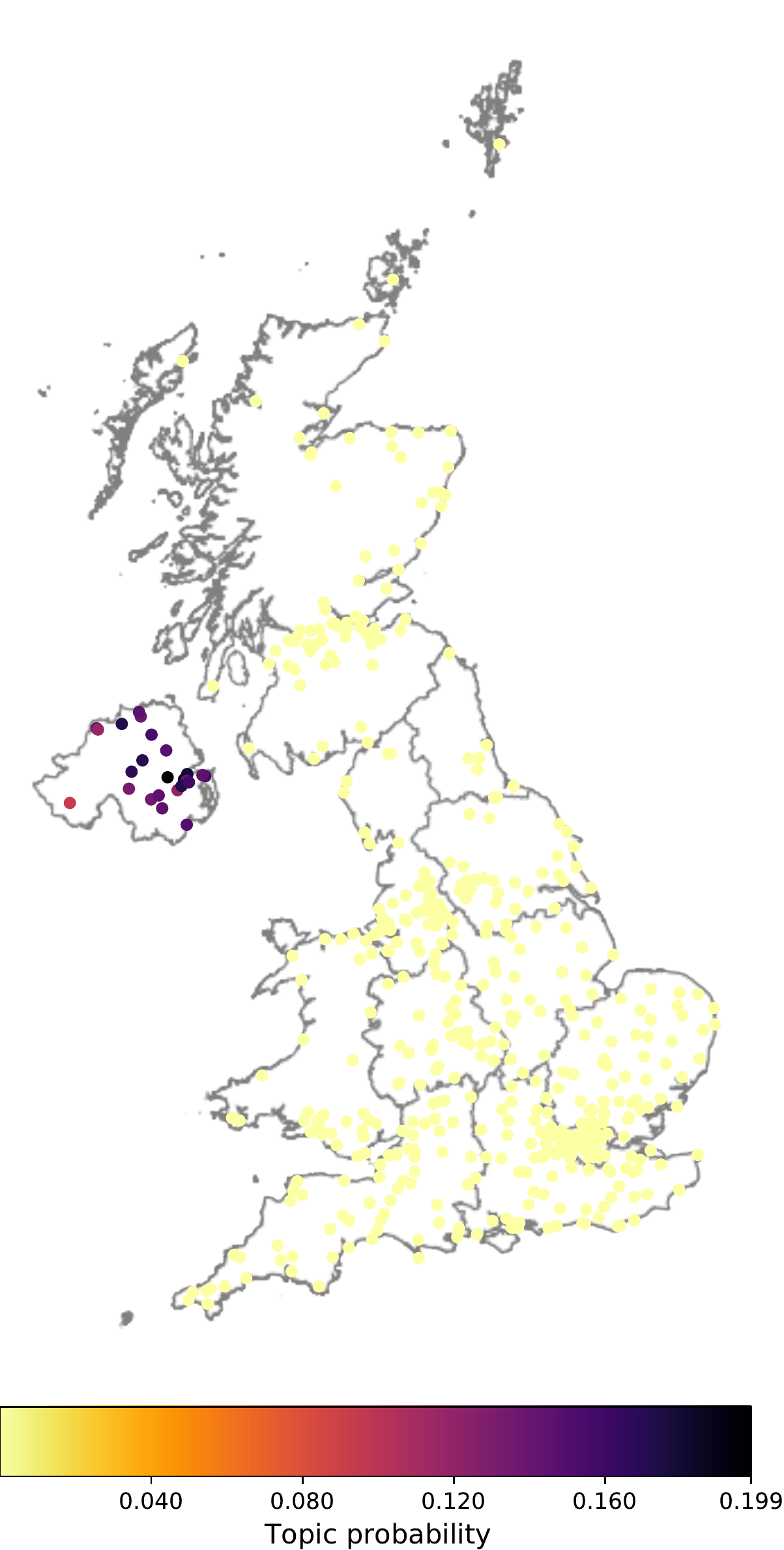}               
            \label{m_2}
        \end{subfigure}     
         \begin{subfigure}[b]{0.32\textwidth}
            \centering 
            \caption{Welsh}
            \includegraphics[width=0.95\textwidth]{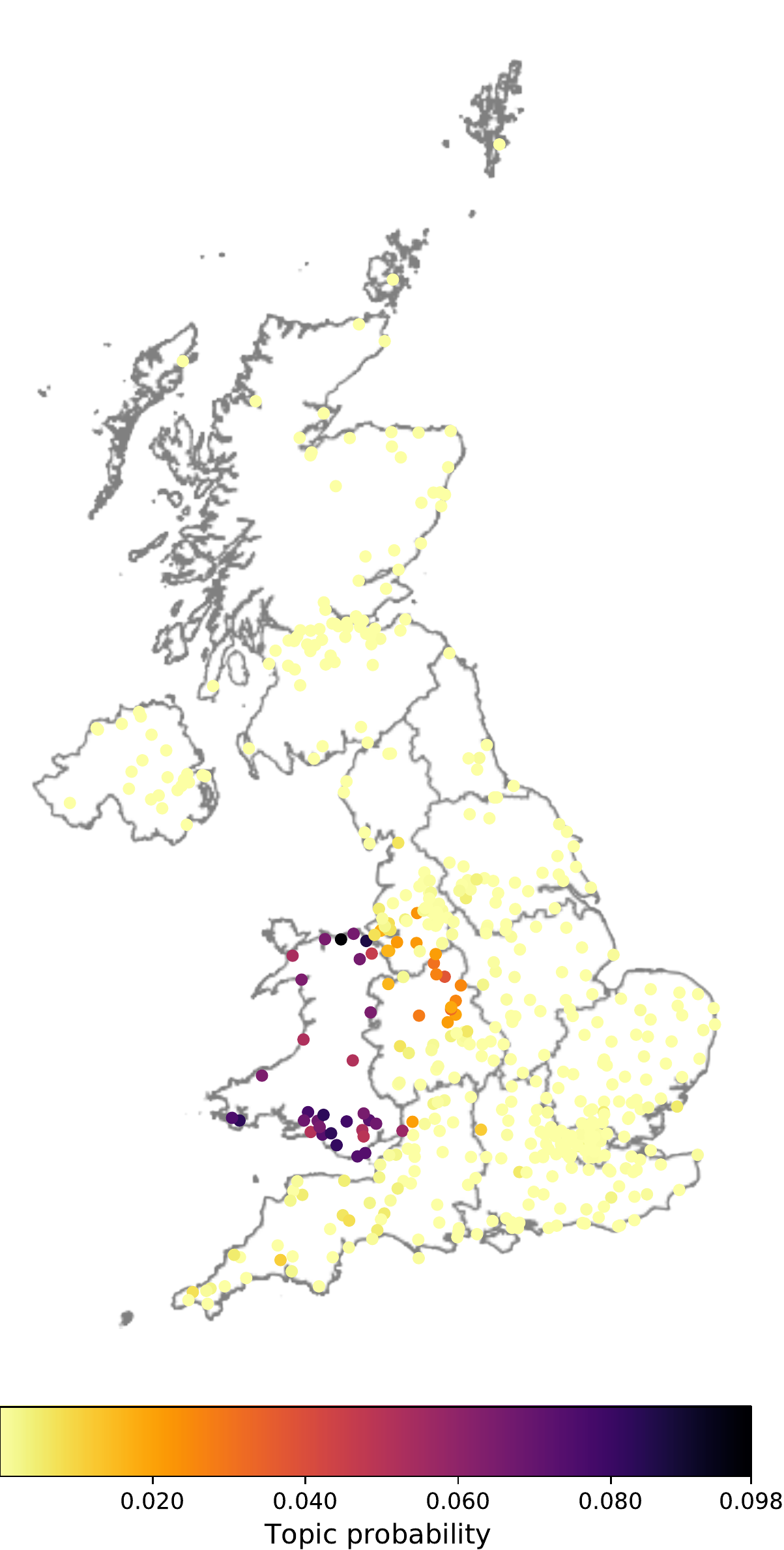}               
            \label{m_3}
             \end{subfigure}         
             \\
        \begin{subfigure}[b]{0.32\textwidth}
            \centering 
            \caption{North and Centre}
            \includegraphics[width=0.95\textwidth]{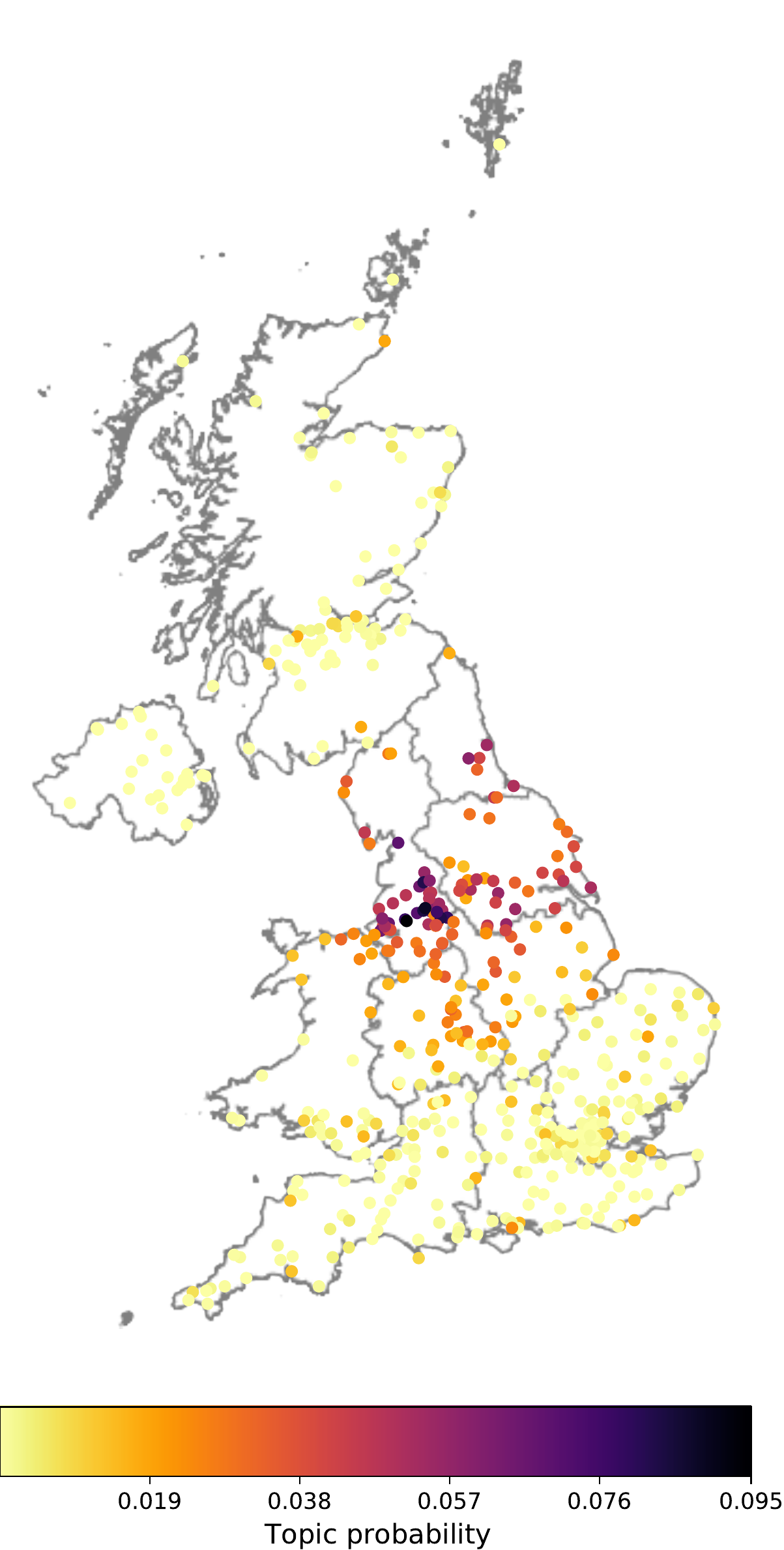}               
            \label{m_4}
        \end{subfigure}    
        \begin{subfigure}[b]{0.32\textwidth}
            \centering 
            \caption{South and Midlands}
            \includegraphics[width=0.95\textwidth]{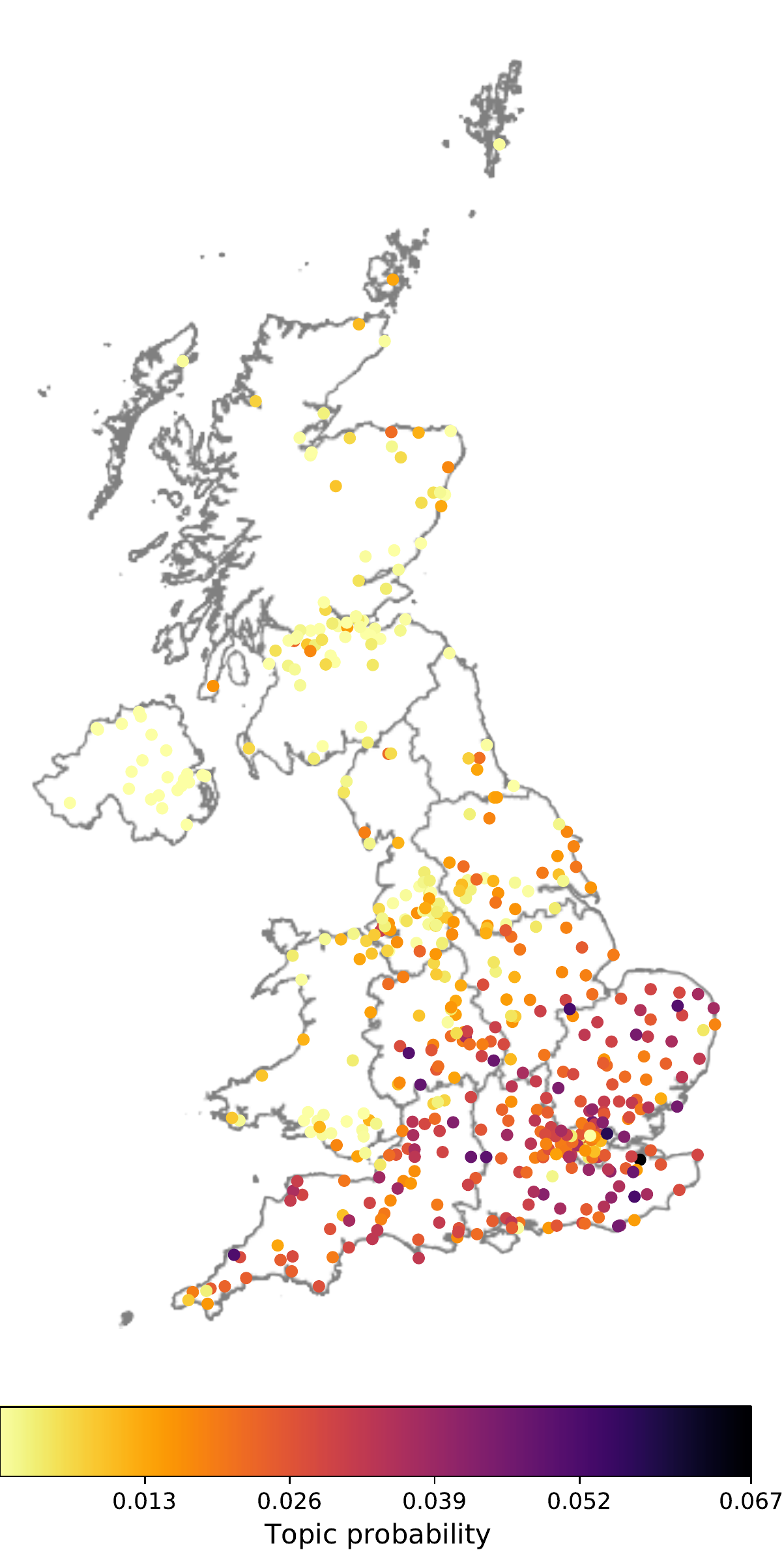}               
            \label{m_5}
        \end{subfigure}
         \begin{subfigure}[b]{0.32\textwidth}
            \centering 
            \caption{Organic}
            \includegraphics[width=0.95\textwidth]{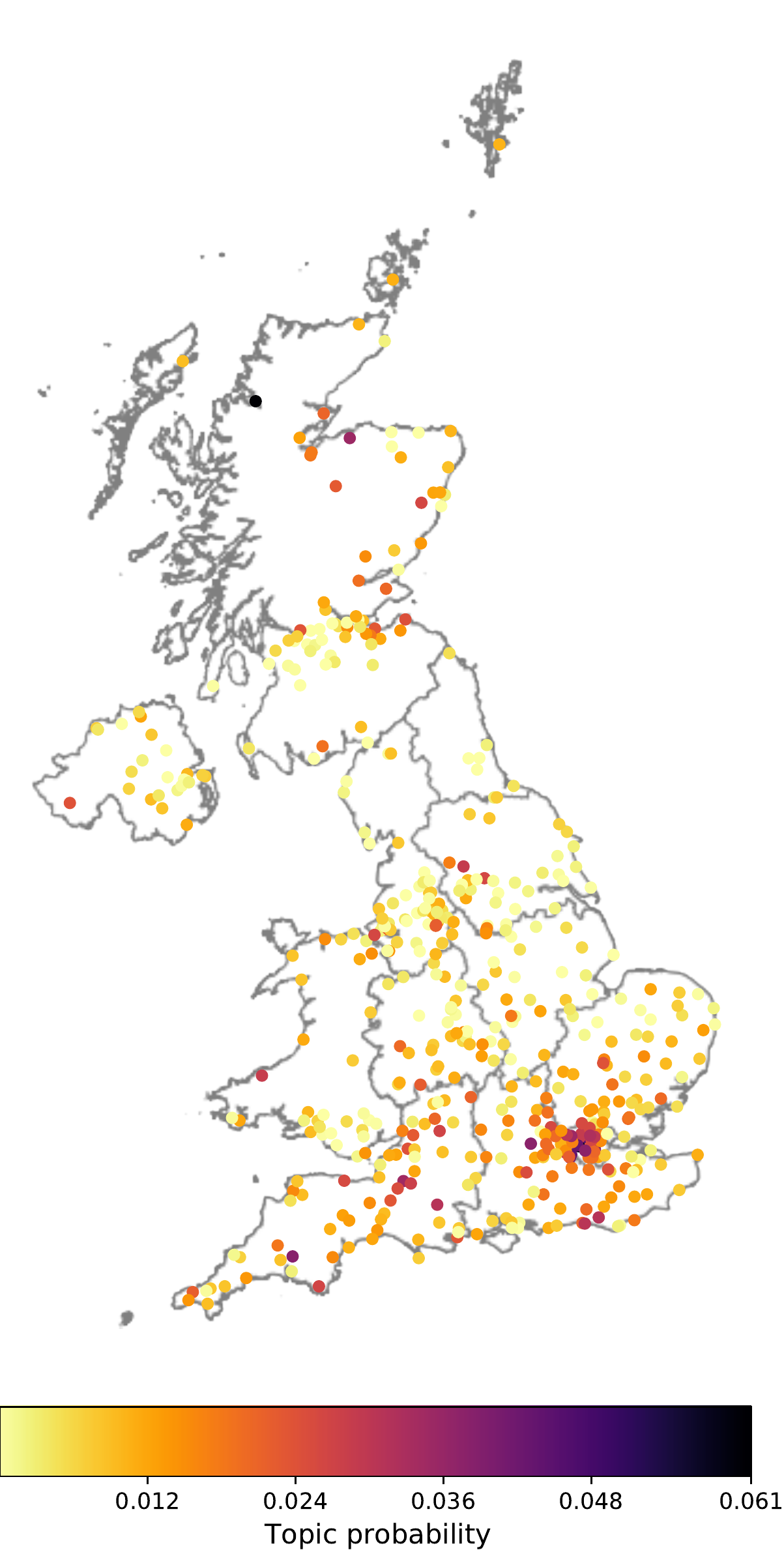}               
            \label{m_6}
        \end{subfigure}      
\caption{Topic probabilities of clustered grocery topics in the UK. Purple and yellow points reflect the largest and smallest topic probabilities, respectively.} 
        \label{maps_uk}
        \end{figure}     
        
\subsection{Mapping topical mixtures}

Interpreting product descriptions may reveal the existence of regional topics, i.e., the Northern Irish/ Scottish/ Welsh topic. Topic interpretations may dismiss regional topics that exhibit products that are not directly linked with specific areas. Thus, we visualise the store-specific topic probabilities at the store's location, aiming to find topics with a regional pattern. We link store postcodes with location coordinates through querying stores' postcodes in the lookup table from the Office for National Statistics \citep{ONS}. Figure \ref{maps_uk} shows the topic probabilities of the six clustered topics mapped across the UK.

Figures \ref{m_1}, \ref{m_2}, \ref{m_3} clearly confirm that the Scottish, Northern Irish and Welsh topics are more likely in their respective constituent countries. More interestingly, Figure \ref{m_3} shows the prevalence of the Welsh topic over neighbouring regions. Figure \ref{m_4} shows high topic probabilities concentrated in the North West and surrounding regions, and Figure \ref{m_5} shows high topic probabilities in the central and southern English regions. We rename both topics as North and Centre and South and Midlands due to their cross-regional prevalence. Figure \ref{m_6}, which maps the Organic topic, shows significant probabilities concentrated in London. 

In comparison to the Scottish, Northern Irish and Welsh topics, interpretations of the most probable topics in the North and Centre, South and Midlands and Organic topics do not easily suggest a geographical pattern. Mapping the store-specific topic probabilities aids the analysis and identification of topics with spatial patterns.
             
\subsection{STM vs LDA}

STM shows two advantages over LDA. Firstly, STM provides topical summaries for stores, by including the store hierarchy above transactions. Secondly, and less obvious, STM discovers topics that are relevant within their store context. In comparison, LDA finds products that are frequently bought together across all transactions. Thus, a product combination that is only frequent in few stores may not be shown among LDA topics. The ability to capture store-specific topics is key to our subsequent spatial modelling analysis.

We compare the 104 STM clustered topics (HC-STM-100) against the posterior summaries of the LDA model with 100 and 200 topics. The posterior summaries of LDA were obtained using the same training data and following the clustering methodology in \cite{vegacarrasco2020modelling}. The posterior summary of LDA with 100 topics (HC-LDA-100) gathers 96 clustered topics and the posterior summary of LDA with 200 topics (HC-LDA-200) gathers 198 clustered topics. 

\begin{figure}[H]
 \centering
 \begin{subfigure}[b]{0.35\textwidth}
            \centering 
            \caption{HC-STM-100 vs HC-LDA-100}
            \includegraphics[width=1\linewidth]{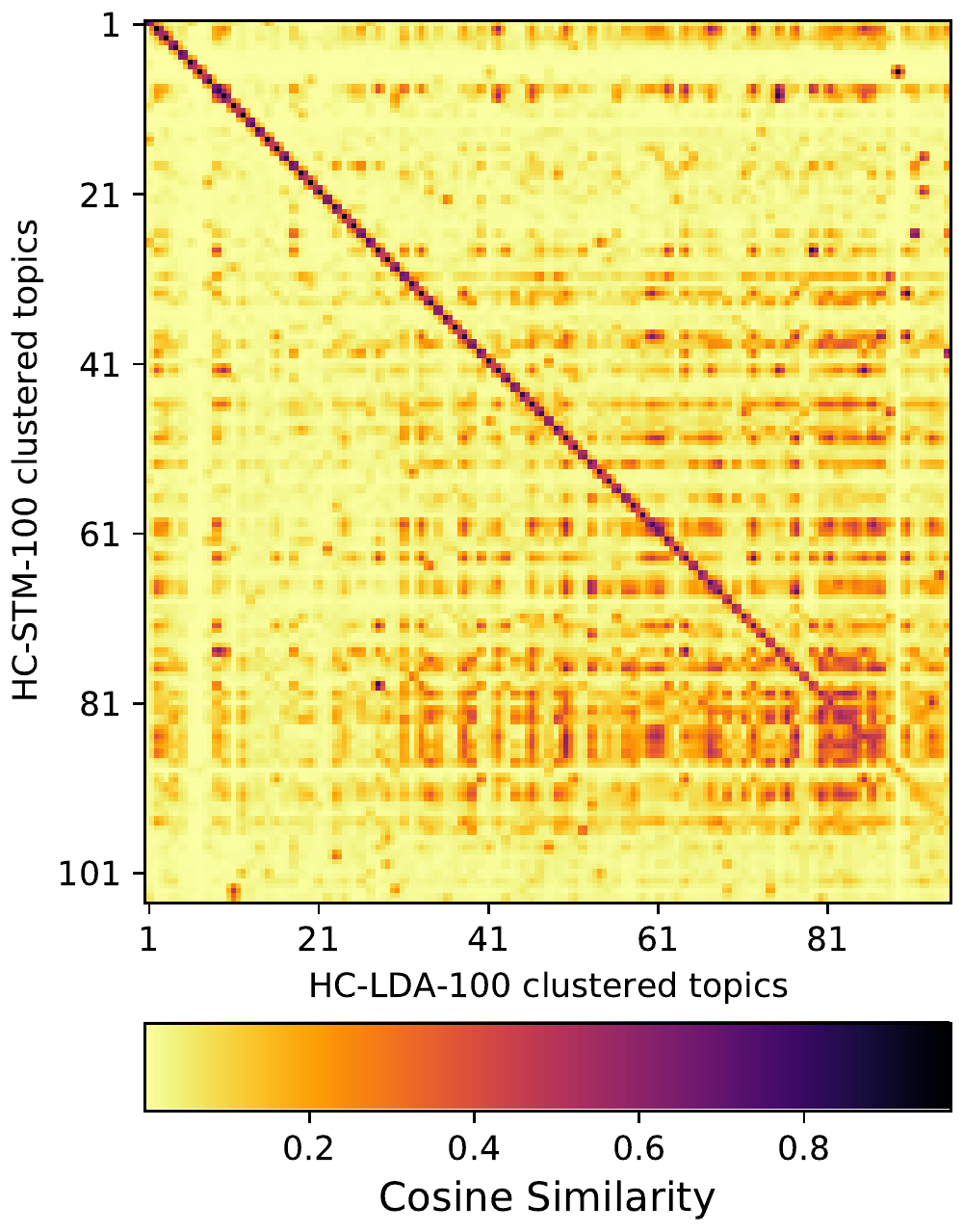}                           
            \label{CD_1}
  \end{subfigure} 
      \begin{subfigure}[b]{0.64\textwidth}
            \centering 
            \caption{HC-STM-100 vs HC-LDA-200}
            \includegraphics[width=1\linewidth]{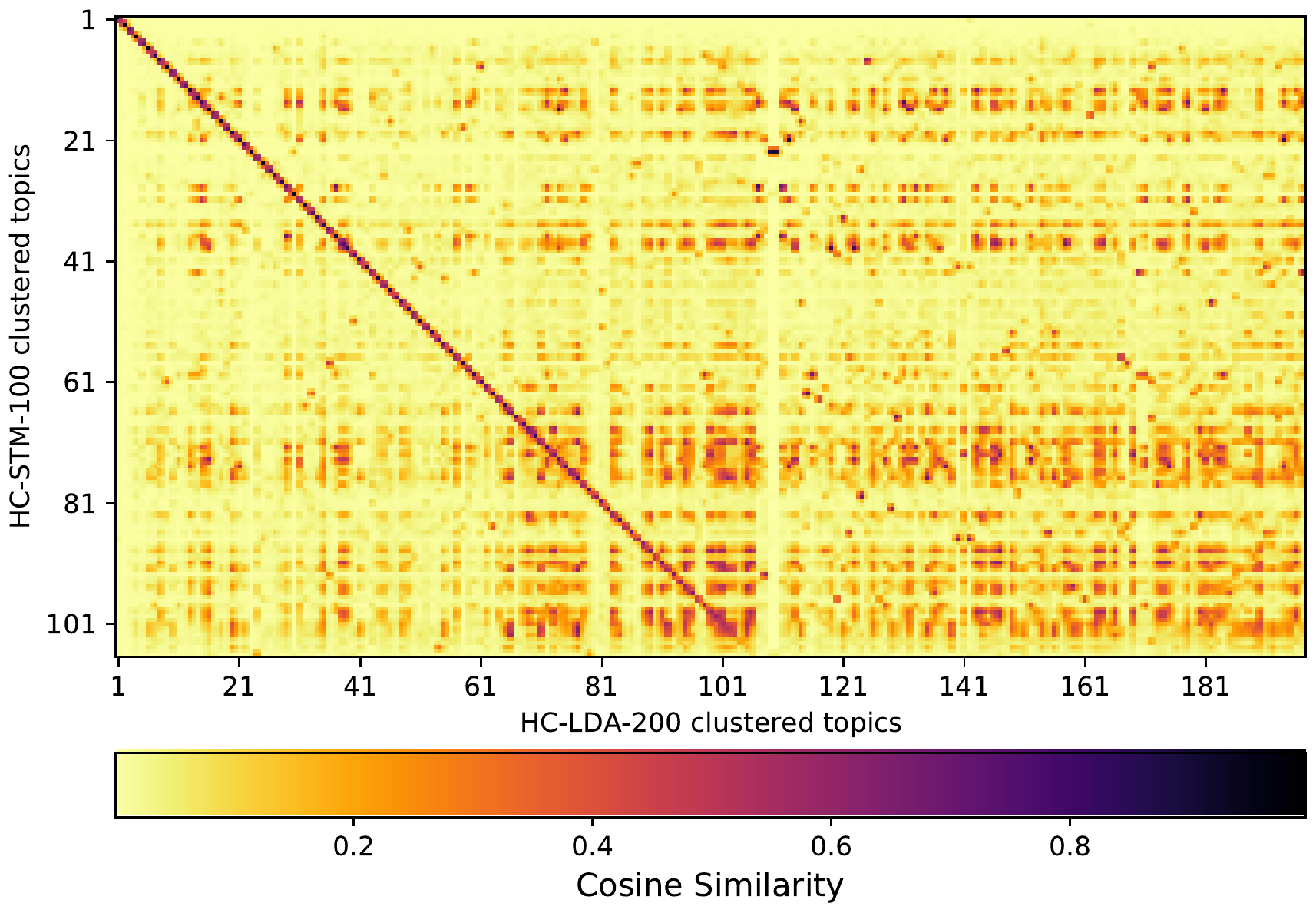}                           
            \label{CD_2}
        \end{subfigure} \\
             \caption{Cosine similarity between clustered topics obtained from posterior summaries of STM with 100 topics and LDA with 100 and 200 topics. Topics have been aligned following a greedy algorithm that at each step searches and pairs topics (that have not been paired) with the highest cosine similarity.}
        \label{Comparison}
\end{figure}

\begin{figure}
 \centering
   \begin{subfigure}[b]{\textwidth}
            \centering 
            \caption{HC-STM-100 vs HC-LDA-100}
            \includegraphics[width=1\linewidth]{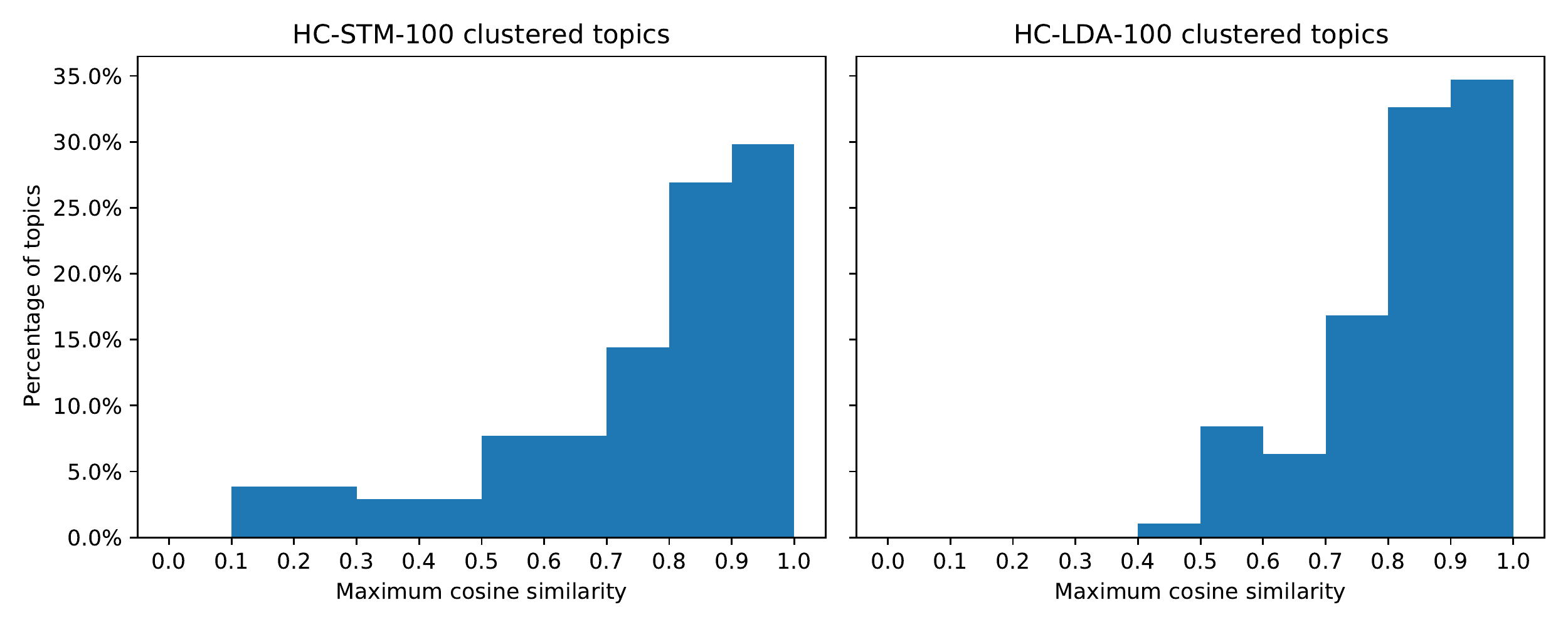}                           
            \label{hist_1}
  \end{subfigure} \\
\begin{subfigure}[b]{\textwidth}
            \centering 
            \caption{HC-STM-100 vs HC-LDA-200}
            \includegraphics[width=1\linewidth]{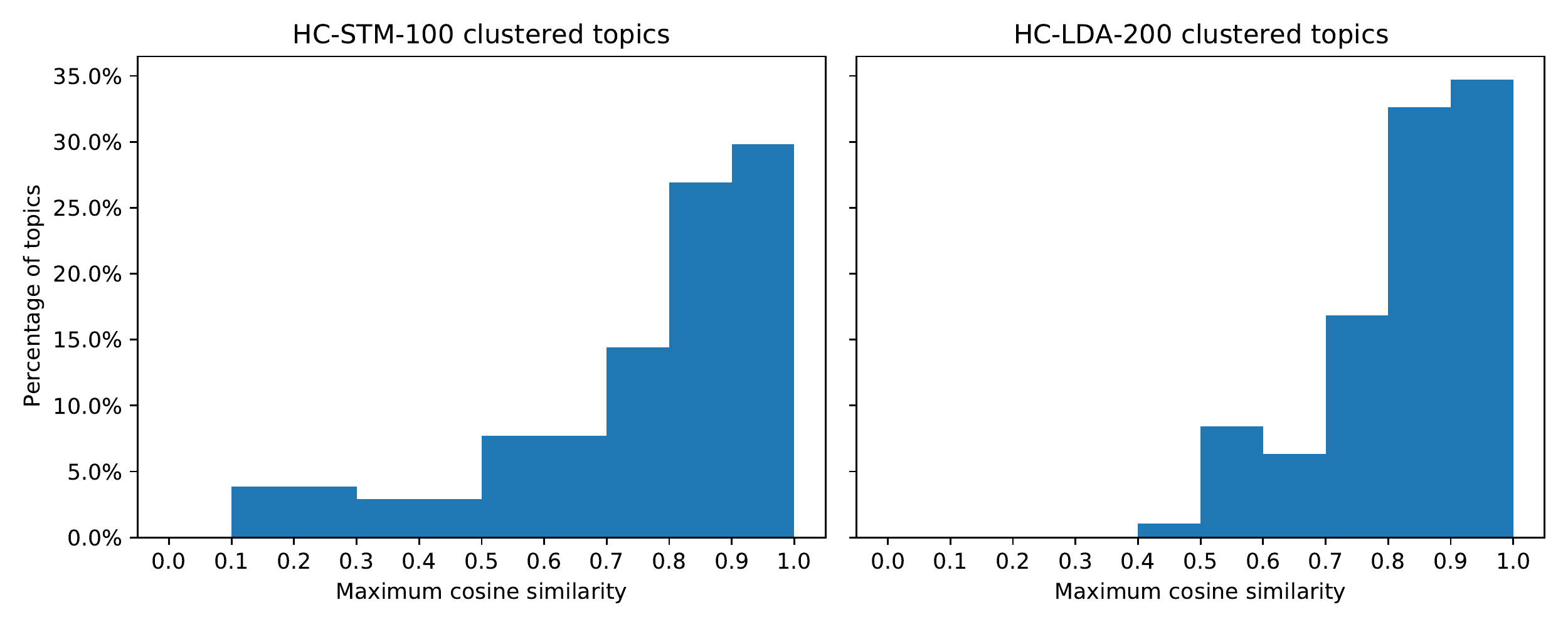}                           
            \label{hist_2}
  \end{subfigure} 
             \caption{Distributions of the maximum cosine distance obtained from each cosine similarity matrix in Figure \ref{Comparison}. Figure \ref{hist_1} plots maximum cosine distances between clustered STM topics (HC-STM-100) against the posterior summary of LDA with 100 topics (HC-LDA-100) (left); and from HC-LDA-100 to HC-STM-100 (right). Figure \ref{hist_2} plots maximum cosine distances between HC-STM-100 against the posterior summary of LDA with 200 topics (HC-LDA-200) (left); and from HC-LDA-200 to HC-STM-100 (right).}
        \label{Comparison}
\end{figure}

\begin{figure}
 \centering
 \begin{subfigure}[b]{1\textwidth}
            \centering 
            \caption{The Northern Irish topic in STM and LDA.}
            \includegraphics[width=1\linewidth]{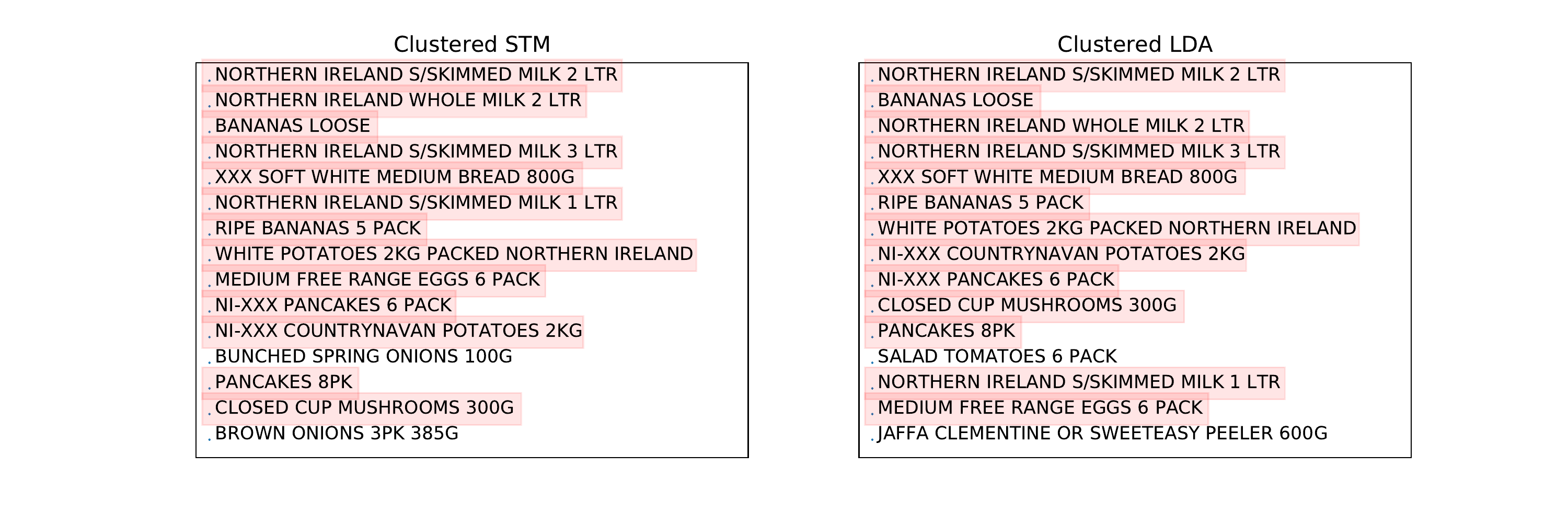}                           
            \label{NI_STM_LDA}
  \end{subfigure} \\
    \begin{subfigure}[b]{1\textwidth}
            \centering 
            \caption{The Welsh topic in STM and its most similar topic in LDA }
            \includegraphics[width=1\linewidth]{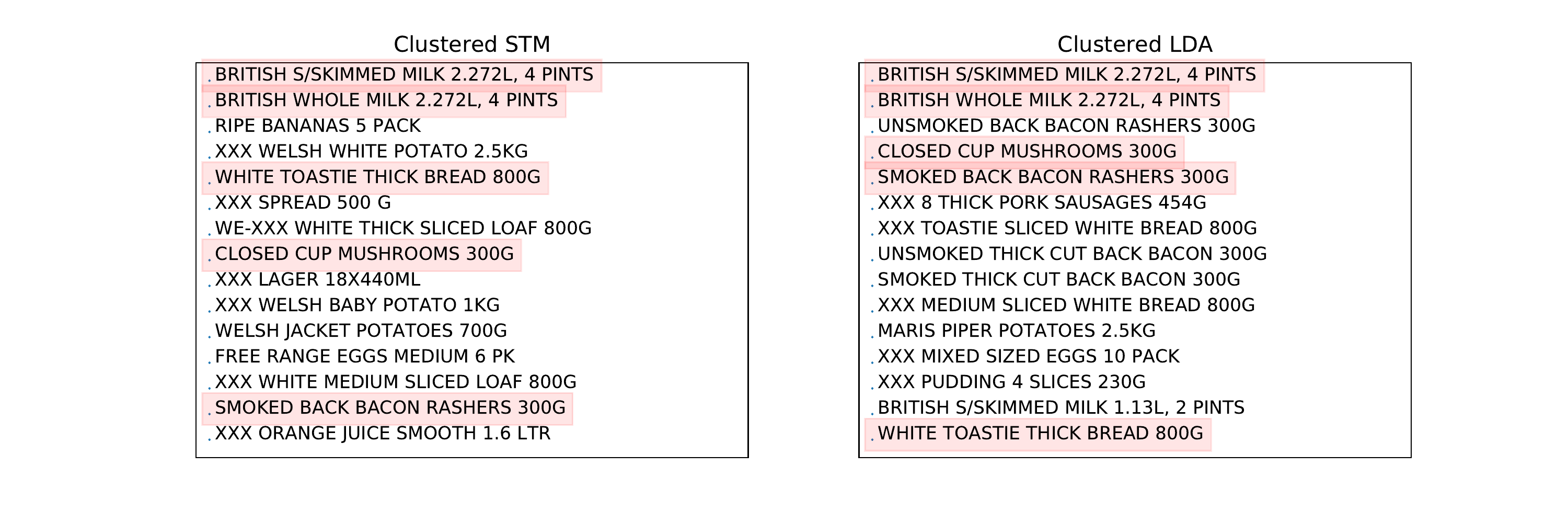}                           
            \label{Welsh_STM_LDA}
        \end{subfigure} 
             \caption{Comparison of topics identified in STM and LDA posterior samples. Highlighted products appear in both topics. While the Northern Irish topic is clearly identified by both models, the Welsh topic is only found by the STM model.}
        \label{TopicComparison}
\end{figure}

Figure \ref{CD_1} shows the cosine similarity between (HC-STM-100) 104 clustered topics and (HC-LDA-100) 96 clustered topics. Clustered topics are ordered to visualise high similarities in the diagonal. As observed, the majority of clustered topics are identified in both models, STM and LDA, with high cosine similarity \( > 0.7\). Figure \ref{hist_1} shows that 70\% of the (HC-STM-100) clustered topics are found among HC-LDA-100 clustered topics, and 85\% of the HC-LDA-100 clustered topics are found among the HC-STM-100 clustered topics with high similarity. For instance, the Northern Irish topic is found in both models with high cosine similarity (0.97). As depicted in Figure \ref{NI_STM_LDA}, Northern Ireland related products rank in the top 15 products in both topics. The Organic topic was also found among HC-LDA-100 clustered topics with high cosine similarity (0.95).

We also compare the 104 STM clustered topics against the 198 LDA clustered topics obtained from summarising LDA posterior samples of 200 topics. This comparison allows identifying the regional topics that were not caught in LDA samples with 100 topics. For instance, the Scottish topic described in Figure \ref{t_2}, is not found in the HC-LDA-100 subset, but it is found in the HC-LDA-200 subset with a cosine similarity of 0.83. As observed in Figure \ref{CD_2}, the majority of the 104 clustered topics are found among the (HC-LDA-200) 198 LDA clustered topics with high cosine similarity (\(> 0.7\)). However, Figure \ref{hist_2} shows that there are still some STM clustered topics that do not match with any of the LDA clustered topic with high similarity. For example, the Welsh topic described in Figure \ref{t_3}, is not found in either of the two subsets of LDA clustered topics. The Welsh topic and the closest clustered topic in HC-LDA-200 (with \( 0.67\) cosine similarity) are listed in Figure \ref{Welsh_STM_LDA}; as observed, few products are shared by the topics but Welsh products are not described in both topics. The North and Centre topic and the South and Midlands topic were not found among the HC-LDA-200 clustered topics either. Perhaps, these regional topics would appear among posterior samples of a larger LDA model, i.e., LDA with 300 topics; however, increasing the model complexity is not only more computationally expensive but also less efficient since topics tend to show less distinctiveness \citep{vegacarrasco2020modelling}.
 
In summary, three out of the six regional topics are identified by STM and the other three regional topics are identified by both STM and LDA models. One of these topics was captured by a larger LDA model. STM shows its strength over LDA by identifying more regional topics.
           
\section{Modeling regional prevelance}\label{spatialAnalysis}

We aim to model topic probabilities across stores in the UK by constructing a linear model with fixed effects associated with the constituent countries of the UK (Wales, Scotland, Northern Ireland) and the nine English regions, and imposing spatial dependency though a Gaussian process that captures residual spatial association as defined in Equation \ref{dataModel}. In this manner, we can quantify the significance of a topic to a region or constituent country. This administrative division was chosen assuming that each country and region would broadly show differences in customer behaviour. Analysis over other subdivisions is possible, but it is out of the scope of this paper.

The dependent variable \(\textbf{Y}_k\) is the \(\textrm{logit}\) transformation of the store-specific \(k^{th}\) topic probabilities \( [\widehat{\theta}_{\textbf{s}_1,k},\widehat{\theta}_{\textbf{s}_2,k},...,\widehat{\theta}_{\textbf{s}_n,k}]\), given by:

\begin{equation}
\label{eq:GP_Y}
 \textbf{Y}_k =\textrm{logit}([\widehat{\theta}_{\textbf{s}_1,k},\widehat{\theta}_{\textbf{s}_2,k},...,\widehat{\theta}_{\textbf{s}_n,k}]),
\end{equation}
where each \(\widehat{\theta}_{\textbf{s}_i,k}\) is the average probability over 30 posterior samples of the \(k^{th}\) topic at store location \(\textbf{s}_i\) from Section \ref{HCSTM}. For simplicity, we assume independence among topic probabilities and model each topic separately, i.e., for each topic, a linear model is constructed. However, topic probabilities of a topical mixture are not independent of each other since they need to sum to 1.


The \(\textrm{logit}\) transformation not only avoids predicting nonsensical values (i.e., topic probabilities \(> 1\) or \(<0\)), but also aids the visualisation of topic probabilities that cannot be appreciated in the original scale. For instance, Figure \ref{GPmodel} (left panel) highlights stores in the South West that are not noticed in Figure \ref{m_3}.

The covariate matrix \(\textbf{X}\) is defined by dummy variables responding to the constituent countries: `North Ireland', `Scotland', `Wales'; and the English regions: `North East', `North West', `Yorkshire and the Humber', `East Midlands', `West Midlands', `South West', `South East', and `East Anglia', where `London' is the reference category.  

The spatial distance \(\textrm{dist}(\textbf{s}_i,\textbf{s}_j)\), which define covariance between stores \(C_{\boldsymbol{\eta}}(\textbf{s}_i,\textbf{s}_j)\), is calculated by firstly finding the latitude-longitude coordinates associated with the store's postcode, secondly computing the distance between pair of coordinates using the Haversine formula \citep{robusto1957cosine}. The Haversine formula provides accurate approximations of distance for locations over large areas. Postcode coordinates are queried from the postcode lookup table from the Office for National Statistics \citep{ONS}. Spatial distance is measured in kilometres. 

We complement the Bayesian hierarchical model with weakly informative priors: \(\sigma^2 \sim \textrm{half}N(0,1)\), \(\beta \sim N(0,10)\); \(\alpha \sim N(0,2)\), and \(\rho \sim IG(2,50)\).

Parameters of the linear Gaussian process regression are estimated with Stan, using 2 MCMC chains which run for 2,000 iterations, 1,000 burn-in iterations, and a thin of five iterations. Convergence of MCMC chains is satisfactory with scale factor reduction \(\widehat{R} = 0.998\). 

\subsection{Prevalence of regional behaviours in the UK}

Table \ref{regional_topics} shows posterior summaries of the linear Gaussian process regression. The intercept can be interpreted as how likely (in logit scale) a topic is at a store in London and vice versa. Positive average coefficients indicate that the topic is more likely than in London. Average coefficients that are highlighted in red correspond to non-zero 95\% credible intervals with \(0>\) upper bound, and bold average coefficients correspond to non-zero 95\% credible intervals with \(0<\) lower bound. 

Unsurprisingly, the Scottish, Northern Irish and Welsh topics show positive average coefficients with non-zero credibility intervals for the respective constituent countries. This indicates that their topic probability largely increases for stores in Scotland, North Ireland and Wales, respectively.

\begin{table}
  \centering  
\scriptsize
\caption{Regression parameters for regional topics \textcolor{red}{Red}/\textbf{bold} mean estimates for coefficients with non-zero credibility intervals that decrease/increase the topic probability, respectively.} 
\begin{tabular}{|c| c c | c c | c c | c c | c c | c c |}
\hline
  \multirow{1}{*}{ }
    & \multicolumn{2}{c|}{Northern}
   & \multicolumn{2}{c|}{Scottish} 
    & \multicolumn{2}{c|}{Welsh}
       & \multicolumn{2}{c|}{North} 
  & \multicolumn{2}{c|}{South}
    & \multicolumn{2}{c|}{Organic}\\
   \multirow{1}{*}{ }
     & \multicolumn{2}{c|}{Irish}
  & \multicolumn{2}{c|}{} 
    & \multicolumn{2}{c|}{}  
    & \multicolumn{2}{c|}{and Centre } 
  & \multicolumn{2}{c|}{and Midlands }
    & \multicolumn{2}{c|}{}\\
    \hline
Parameter &	Avg. & SE &	Avg. & SE & 	Avg. & SE &	Avg. & SE &	Avg. & SE & 	Avg. & SE\\
\hline
Intercept &	\textcolor{red}{-10.4}&	0.02&	\textcolor{red}{-9.52}&	0.03&	\textcolor{red}{-8.9}&	0.04&	\textcolor{red}{-6.34}&	0.04&	\textcolor{red}{-4.42}&	0.02&	\textcolor{red}{-4.62}&	0.05\\
Northern Ireland &	  \textbf{8.67}&	0.03&	-0.72&	0.04&	-1.44&	0.07&	\textcolor{red}{-4.11}&	0.05&	\textcolor{red}{-5.77}&	0.03&	-1.25&	0.06\\
Scotland &	0.19&	0.02&	\textbf{6.84}&	0.04&	-1.12&	0.05&	\textcolor{red}{-1.93}&	0.04&	\textcolor{red}{-1.82}&	0.03&	-1.34&	0.06\\
Wales &	-0.4&	0.03&	-0.57&	0.03&	\textbf{5.63}&	0.07&	0.39&	0.04&	\textcolor{red}{-2.27}&	0.03&	-1.27&	0.06\\
North West &	0.15&	0.03&	\textbf{1.54}&	0.04&	0.1&	0.08&	\textbf{3.3}&	0.05&	-0.99&	0.03&	-1.91&	0.06\\
North East &	-0.86&	0.04&	\textbf{3.27}&	0.05&	-0.33&	0.08&	\textbf{3.05}&	0.06&	-1.25&	0.04&	-2.5&	0.07\\
Yorkshire &	0.04&	0.03&	1.08&	0.04&	-0.43&	0.05&	\textbf{2.98}&	0.06&	-0.43&	0.03&	-1.68&	0.05\\
West Midlands &	-0.15&	0.02&	-0.24&	0.03&	\textbf{1.89}&	0.07&	\textbf{1.95}&	0.05&	0.26&	0.03&	-1.01&	0.05\\
East Midlands &	-0.47&	0.03&	0.68&	0.04&	0.67&	0.05&	1.45&	0.05&	0.31&	0.06&	-1.47&	0.05\\
East Anglia &	-0.27&	0.02&	-0.28&	0.03&	-0.38&	0.05&	-0.31&	0.04&	\textbf{0.99}&	0.02&	-1.03&	0.05\\
South East &	-0.21&	0.2&	0.56&	0.03&	-0.25&	0.04&	-1.07&	0.04&	0.66&	0.02&	-0.51&	0.05\\
South West &	-0.26&	0.2&	-0.1&	0.03&	\textbf{1.26}&	0.05&	-0.64&	0.04&	0.71&	0.03&	-0.02&	0.05\\
\hline
Length-scale \(\rho\) &	63.85&	5.95&	92.07&	19.95&	55.31&	1.32&	51.32&	15.53&	50.23&	3.84&	34.67&	3.13\\
Amplitude \(\alpha\) &	0.13&	0.01&	0.3&	0.03&	1.04&	0.01&	0.74&	0.02&	0.23&	0.01&	0.86&	0.02\\
\(\sigma\) &	0.78&	0.01&	1.38&	0.01&	1.43&	0.01&	1.37&	0.01&	1.15&	0.01&	1.58&	0.01\\
 \hline
\end{tabular}
\label{regional_topics} 
\end{table}

\begin{figure}
            \centering 
            \includegraphics[width=1\linewidth]{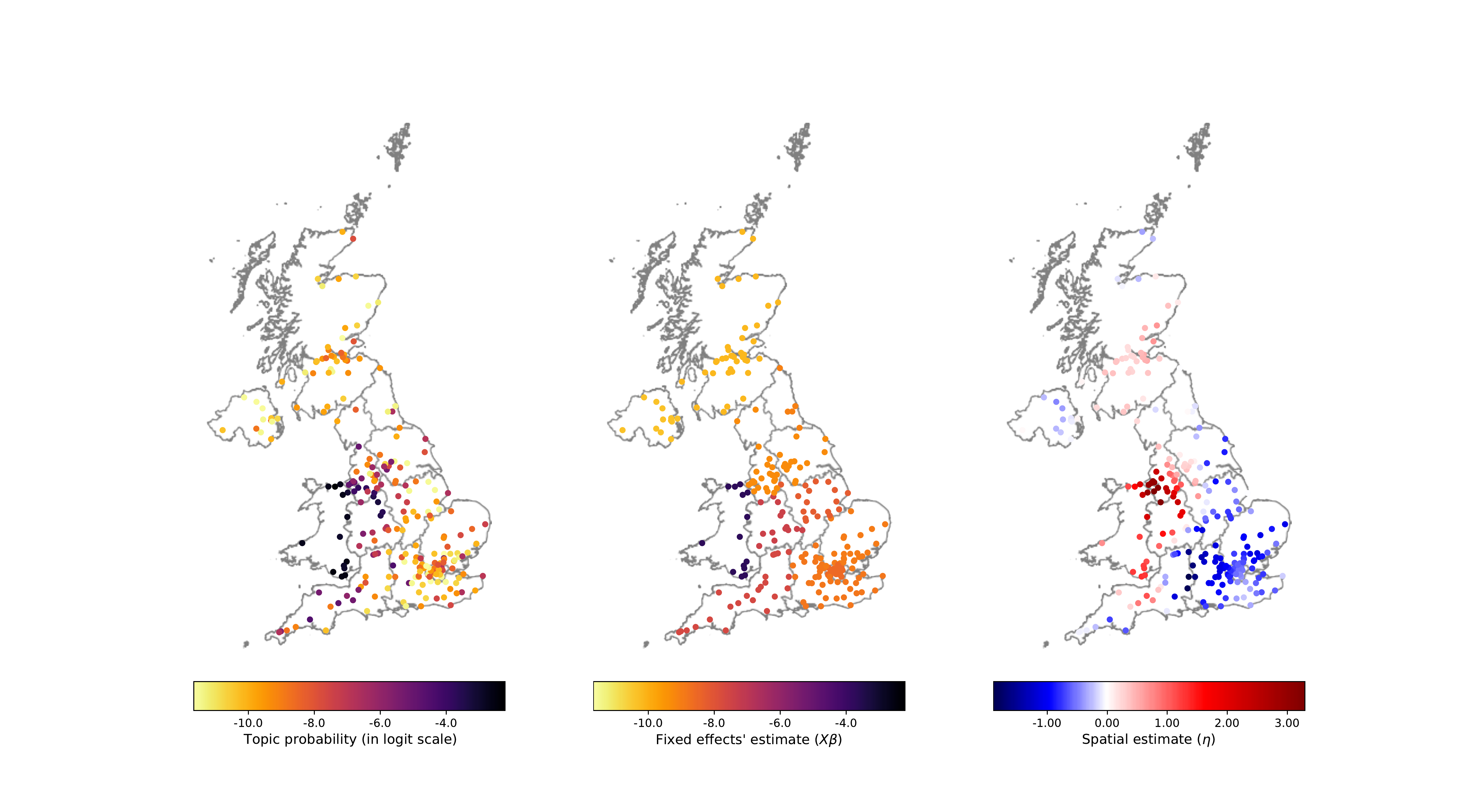}                           
             \caption{Welsh topic: (left panel) observed topic probabilities in logit scale; (central panel) probability estimates (in logit scale) using only fixed effects; (right panel) spatial residuals captured by the Gaussian process.}
        \label{GPmodel}
\end{figure}

Interestingly, Wales's and Scotland's neighbouring regions show positive average coefficients with non-zero credibility intervals, i.e., North East and North West to the Scottish topic and West Midlands and South West to the Welsh topic. As shown in Figure \ref{GPmodel} (central panel), probability estimates (in logit scale) of the Welsh topic for stores in West Midlands and South West are greater than the probability estimates of the Welsh topic at stores in further regions. Moreover, the Gaussian process captures spatial residual distinguishing the stores in the neighbouring regions that are close to Wales from the stores (in the same regions) that are at further distances, as demonstrated in Figure \ref{GPmodel} (right panel).

The coefficients for the North and Centre topic clearly show that the topic is more likely in the North West, North East, Yorkshire and West Midlands and is less likely in Northern Ireland and Scotland. On the other hand, the coefficients for the South and Midlands show that on average the topic is more likely in the southern and central English regions; however, only the coefficient of East England has a non-zero 95\% credibility interval. 

The Organic topic shows a different pattern, its average coefficients are negative; this indicates that the probability of the Organic topic is on average lower than the average topic probability in London. In other words, the Organic topic is more likely in London than in any other region or constituent country; however, the coefficients show 95\% credible intervals containing zero, suggesting that the regional effect may not be significant.

\begin{figure}
 \centering
 \begin{subfigure}[b]{0.48\textwidth}
            \centering 
            \caption{Welsh topic}
            \includegraphics[width=1\linewidth]{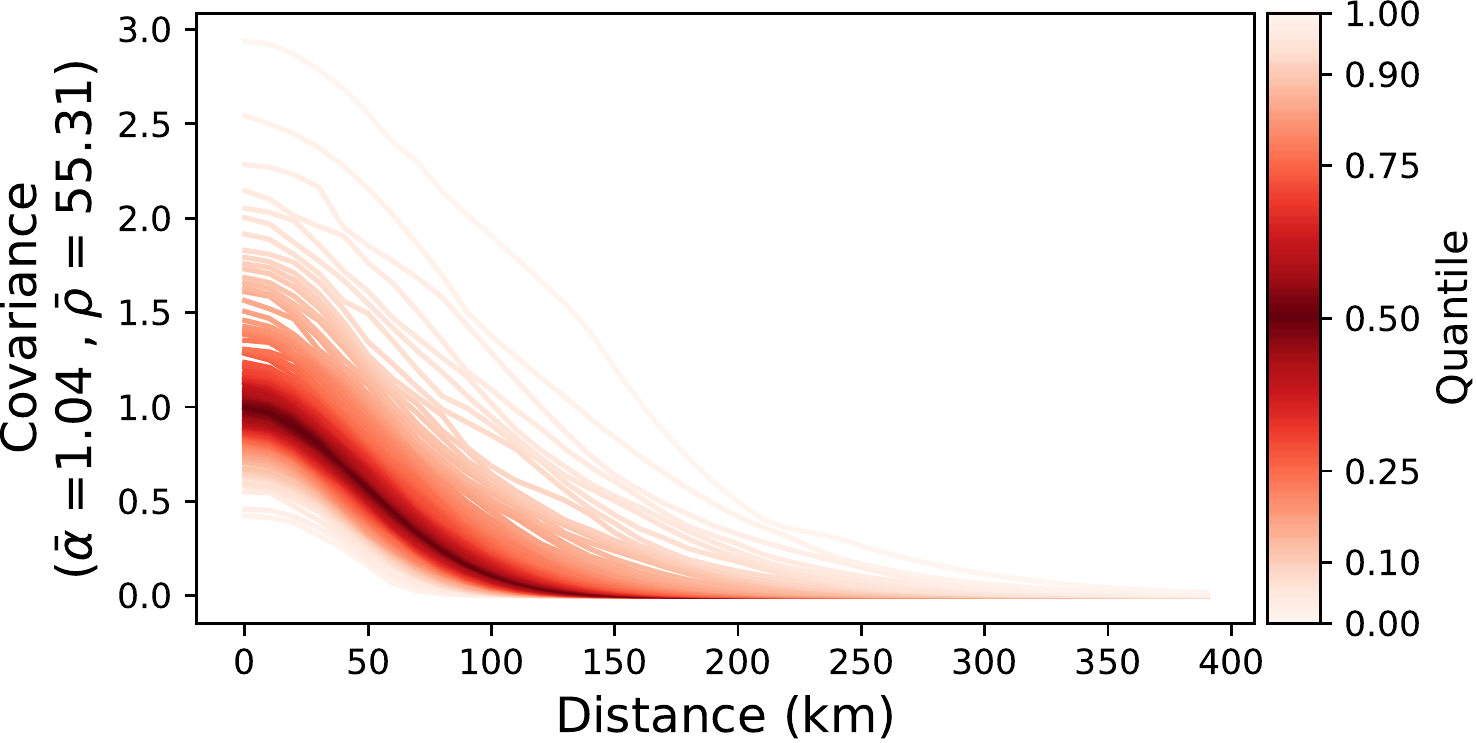}                           
            \label{cov_1}
  \end{subfigure} 
   \begin{subfigure}[b]{0.48\textwidth}
            \centering 
            \caption{Organic topic}
            \includegraphics[width=1\linewidth]{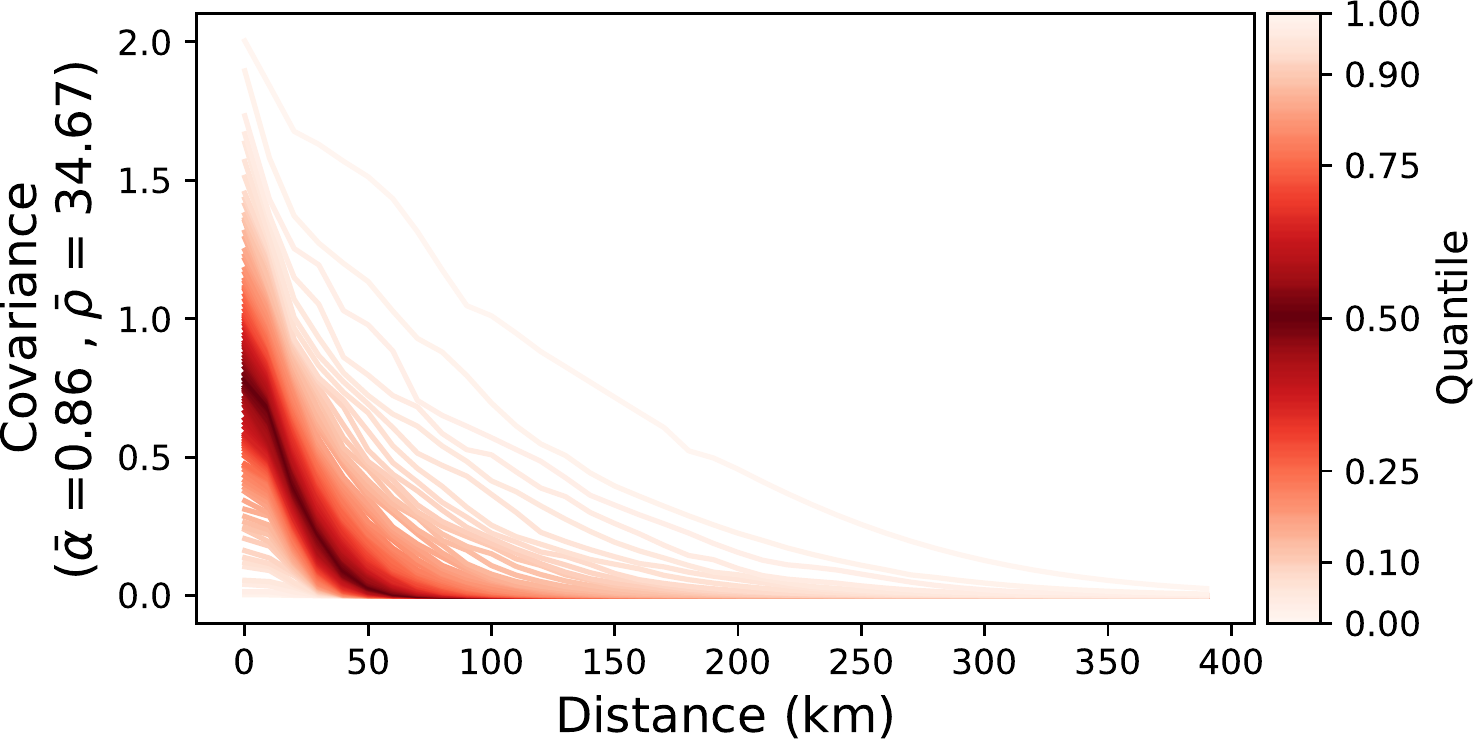}                           
            \label{cov_2}
  \end{subfigure} 
  \caption{Covariance function of the Welsh topic and Organic topic. Lines are computed with posterior samples of \(\alpha\) and \(\rho\).}
\label{covariancefunction}
\end{figure}

The covariance parameters length-scale \(\rho\) and amplitude \(\alpha\) model the covariance between stores, which is stronger when the spatial distance is smaller than \(\rho\) and when \(\alpha\) is significantly larger from zero. The Welsh and the Organic topic show strong covariance as depicted in Figure \ref{covariancefunction}. On the other hand, the Northern Irish topic and the Scottish topic show small values of \(\alpha\) indicating weak covariance functions.

\subsection{Linear Gaussian process regression vs Linear regression}

Here, we compare \textit{mean squared error} and the log of the probability density on held-out data obtained from model topic prevalence using linear Gaussian process regression (LGPR) and the linear regression (LR). We will show that the former model retrieves more accurate estimates and better predictive likelihood by modelling residual spatial effect. 

Table \ref{mse} shows that LGPR improves the prediction of topic probabilities of the Welsh, English-Northern and Centre, South and Midlands and Organic topics. The difference between the mean squared error of these topics is statistically significant at the 0.05 level, indicating that the Gaussian process provides significant model improvement. Similarly, the log predictive likelihood of the four aforementioned topics is significantly better at the 0.05 level. On the contrary, the LGPR doesn't show significantly improved predictions of the Scottish and Northern Irish topics. The difference of their mean squared errors is not statistically significant at the 0.05 level; however, the LGPR shows significantly better predictive log-likelihood at the 0.05 level.

\begin{table}[H]
  \centering  
\scriptsize
\caption{Comparison of the linear Gaussian process regression (LGPR) vs linear regression (LR). lppd: log posterior predictive density on test data. p-values are computed for the pointwise difference of the two methods at each observation in the test set.} 
\begin{tabular}{| c | c | c | c | c | c | c |}
\hline
& Northern & Scottish & Welsh & English-North & English-South & Organic\\
& Irish  &  &  & and Centre  &  and Midlands  & \\  
\hline
LR: MSE (SE)&	0.64 (0.001)&	2.16 (0.004)&	3.18 (0.006)&	3.36 (0.007)&	1.65 (0.004)&	3.39 (0.007)\\						
LGPR: MSE (SE)&	0.63 (0.001)&	2.15 (0.004)&	2.64 (0.005)&	3.24 (0.004)&	1.62 (0.003)&	3.18 (0.006)\\						
p-value  &	0.5877 &	0.1664 &	0.0000 &	0.0000 &	0.0000 &	0.0000\\
\hline						
LR lppd (SE)&	-298.3 (0.30)&	-450.2 (0.25)&	-499.1 (0.23)&	-513.5 (0.31)&	-418.8 (0.38)&	-506 (0.26)\\					
LGPR lppd (SE)&	-296.5 (0.28)&	-449.3 (0.26)&	-476.9 (0.26)&	-504.9 (0.43)&	-412.6 (0.40)&	-493.9 (0.27)\\						
p-value &	0.0000&	0.0169&	0.0000&	0.0000&	0.0000&	0.0000\\
 \hline
\end{tabular}
\label{mse} 
\end{table}

\begin{figure}[H]
 \centering
 \begin{subfigure}[b]{0.32\textwidth}
            \centering 
            \caption{Welsh topic}
            \includegraphics[width=0.7\linewidth]{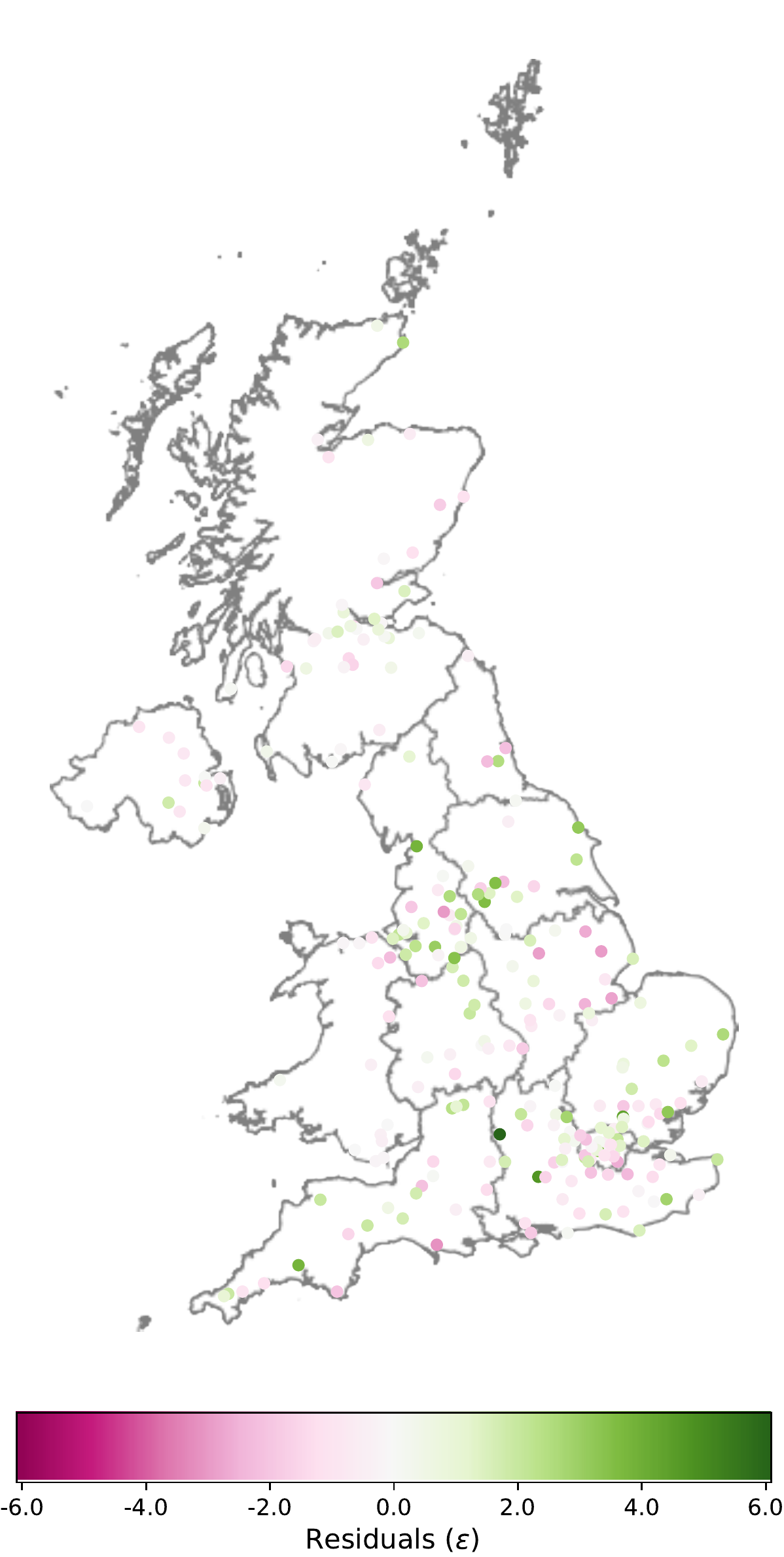}                           
            \label{gpre_1}
  \end{subfigure} 
   \begin{subfigure}[b]{0.31\textwidth}
            \centering 
            \caption{North and Centre topic}
            \includegraphics[width=0.7\linewidth]{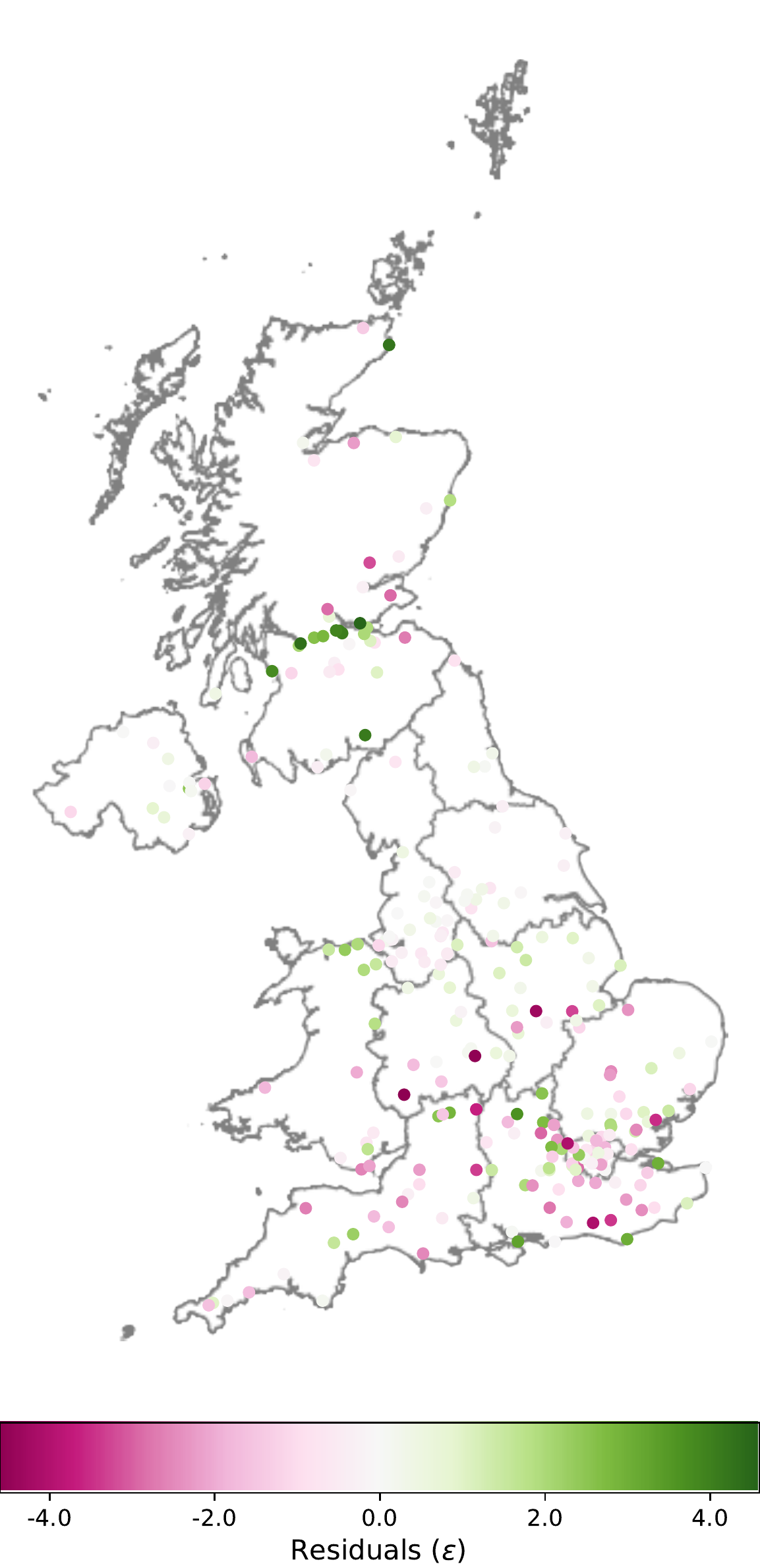}                           
            \label{gpre_2}
  \end{subfigure} 
  \begin{subfigure}[b]{0.31\textwidth}
            \centering 
            \caption{Organic topic}
            \includegraphics[width=0.7\linewidth]{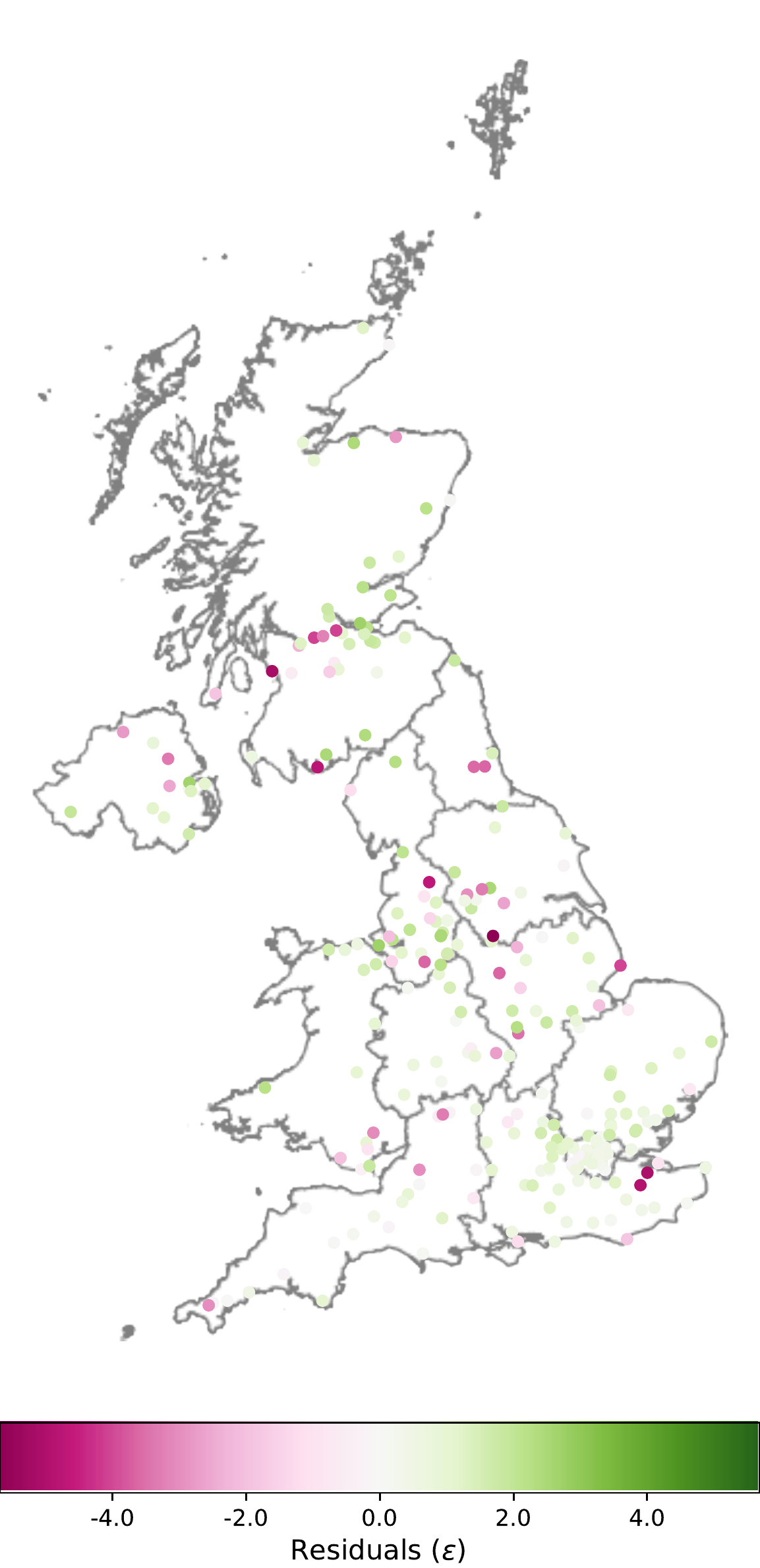}                           
            \label{gpre_3}
  \end{subfigure} 
  \caption{Residuals of modelling the Northern Irish/ North and Centre/ Organic topic with LGPR.}
\label{residuals}
\end{figure}

Examining LGPR residuals in Figure \ref{residuals}, we still observe spatial patterns that are not captured by the Gaussian process. For example, concentrations of underestimated probabilities around North West in Figure \ref{gpre_1}, around the centre of Scotland in Figure \ref{gpre_2}, around South West and East Anglia in Figure \ref{gpre_3}; and overestimated probabilities around South East in Figure \ref{gpre_2}. Further work could explore the Gaussian process with non-stationary covariance to capture local spatial patterns. 

\section{Conclusions}\label{conclusions}

In this paper, we showed that STM is powerful in the analysis of transaction retail data, identifying topics that characterise various customer needs, particularly, those that reflect regional demand. STM harnesses store structure, describing transactions and stores as topical mixtures. More importantly, STM can identify regional topics that otherwise would be overseen by the widely used topic model, the LDA. Aggregating multiple samples of the posterior distribution and selecting topic modes allow the identification of certain and meaningful topics, achieving better data representations and capturing posterior variability. Topic analysis, through LGPR, quantifies regional effects and captures spatial dependence through the squared exponential covariance function. Further work could explore the analysis of spatial topics using other geographical hierarchies such as middle layer super output areas (MSOAs) and non-stationary models such as the non-stationary Gaussian process in \cite{heinonen2016non}, which may capture local spatial residuals. 

\clearpage
\begin{appendix}

\section{MCMC convergence of STM with 100 topics}
\label{convergence_STM}

We evaluate four Markov chains of STM with 100 topics. Markov chains are run for 100,000 iterations with a burn-in period of 80,000 iterations. Log-likelihood is measured at every 10 iterations. We calculate the potential scale reduction factor using 8,000 samples. 

\begin{figure}[H]
    \centering
    \includegraphics[width=1\linewidth]{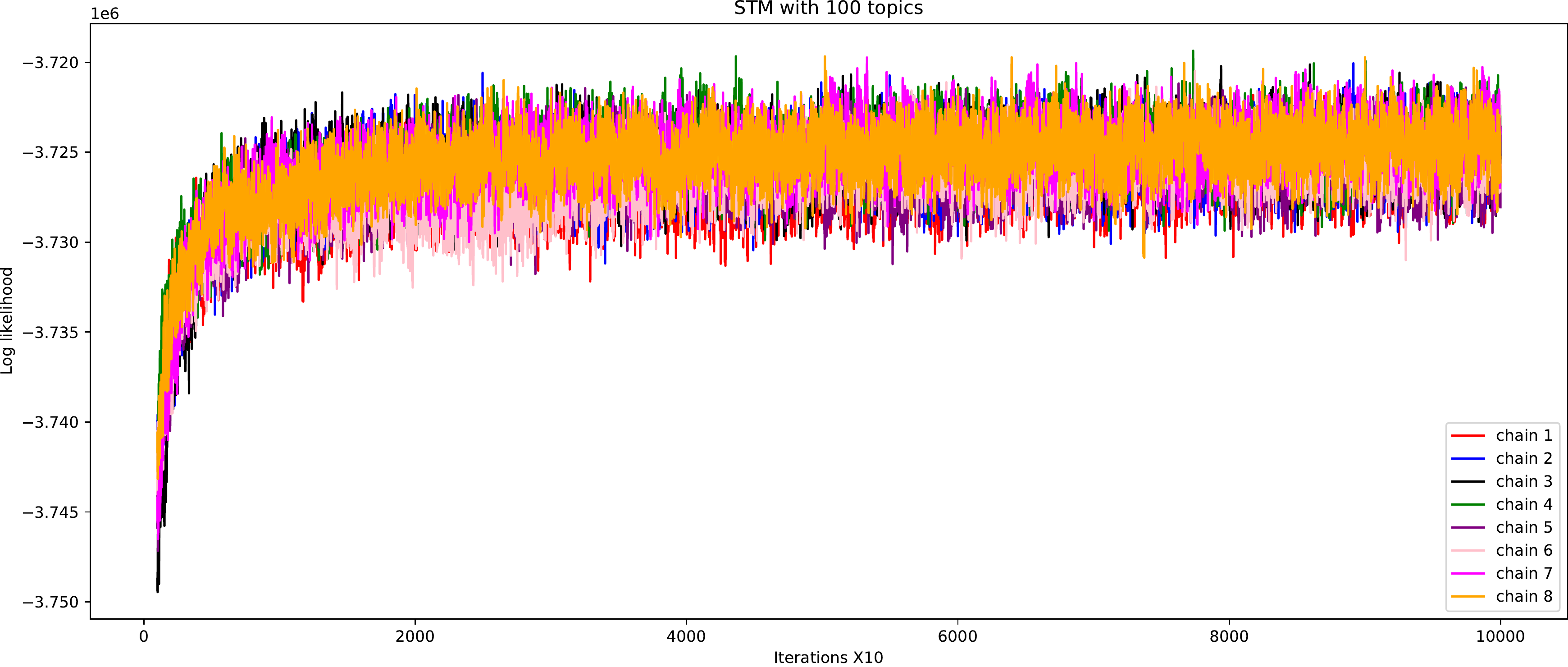}
  \caption{Markov Chains of STM with 100. Potential scale reduction factor \(\hat{R}: 1.07\).}
  \label{mcmc_STM}
    \end{figure}

\section{MCMC convergence of Clustered STM topics}
\label{convergence_HCSTM}
    
We run STM with 104 clustered topics known a priori for 1,500 iterations and burn-in period of 1,000 iterations.
    
\begin{figure}[H]
    \centering
    \includegraphics[width=1\linewidth]{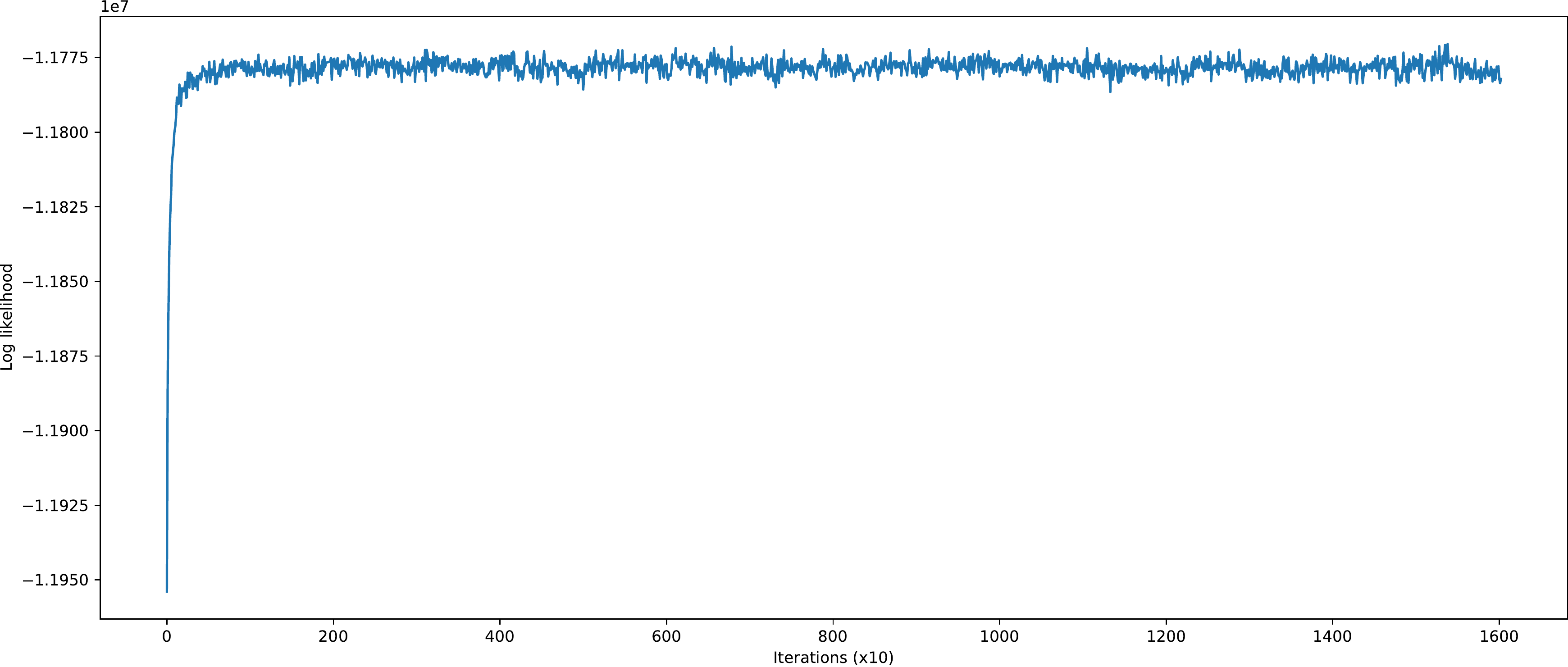}
  \caption{Markov Chain of STM with 104 clustered topics known a priori.  Potential scale reduction factor \(\hat{R}: 0.998\).}
  \label{mcmc_HCSTM}
    \end{figure}
        
\section{Clustering of STM topics.}
\label{Posterior_STM}

\begin{figure}[H]
 \centering
 \begin{subfigure}[b]{0.48\textwidth}
            \centering 
            \caption{Generalisation}
            \includegraphics[width=1\linewidth]{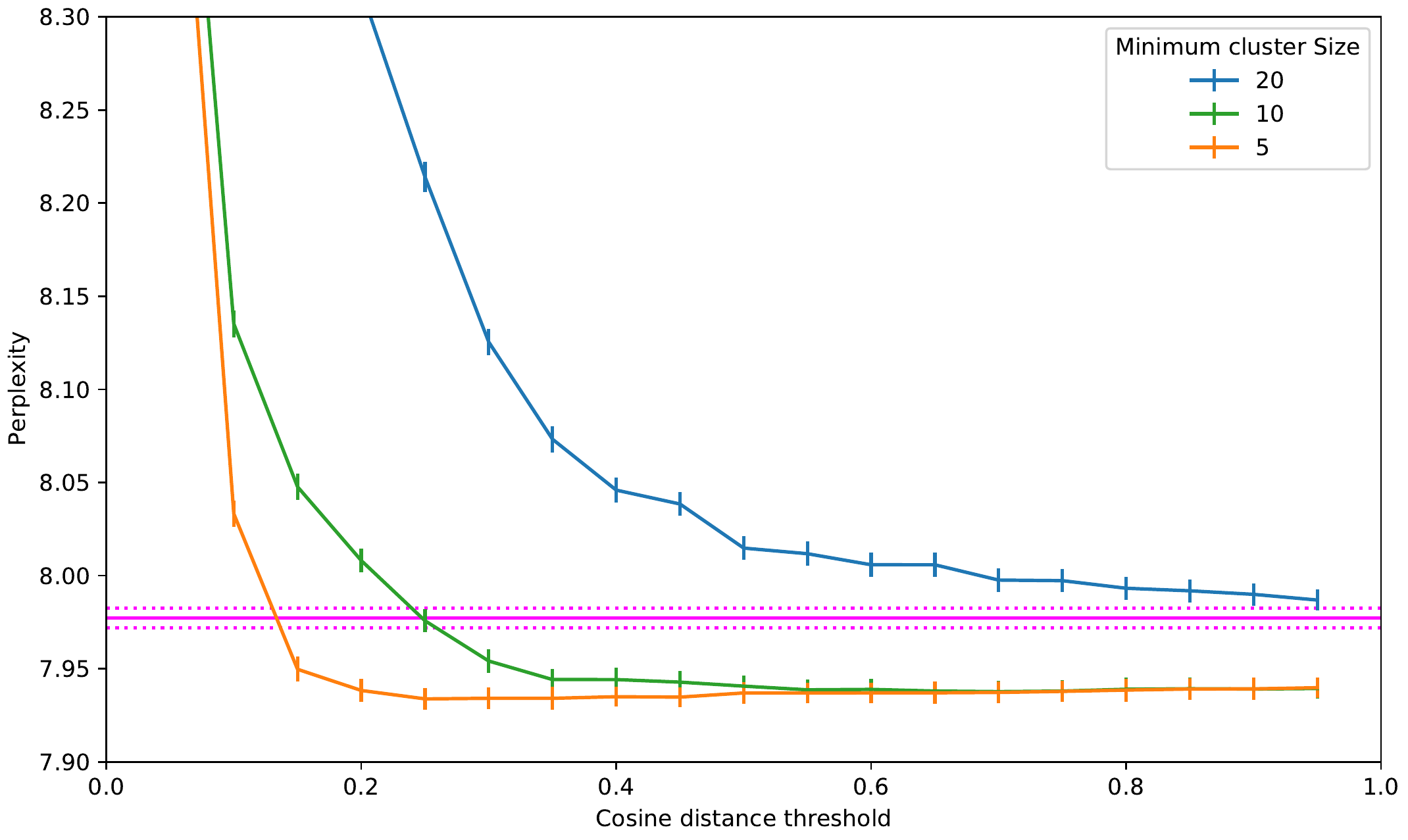}                           
            \label{Performance_1}
  \end{subfigure} 
  \begin{subfigure}[b]{0.48\textwidth}
            \centering 
            \caption{Coherence}
            \includegraphics[width=1\linewidth]{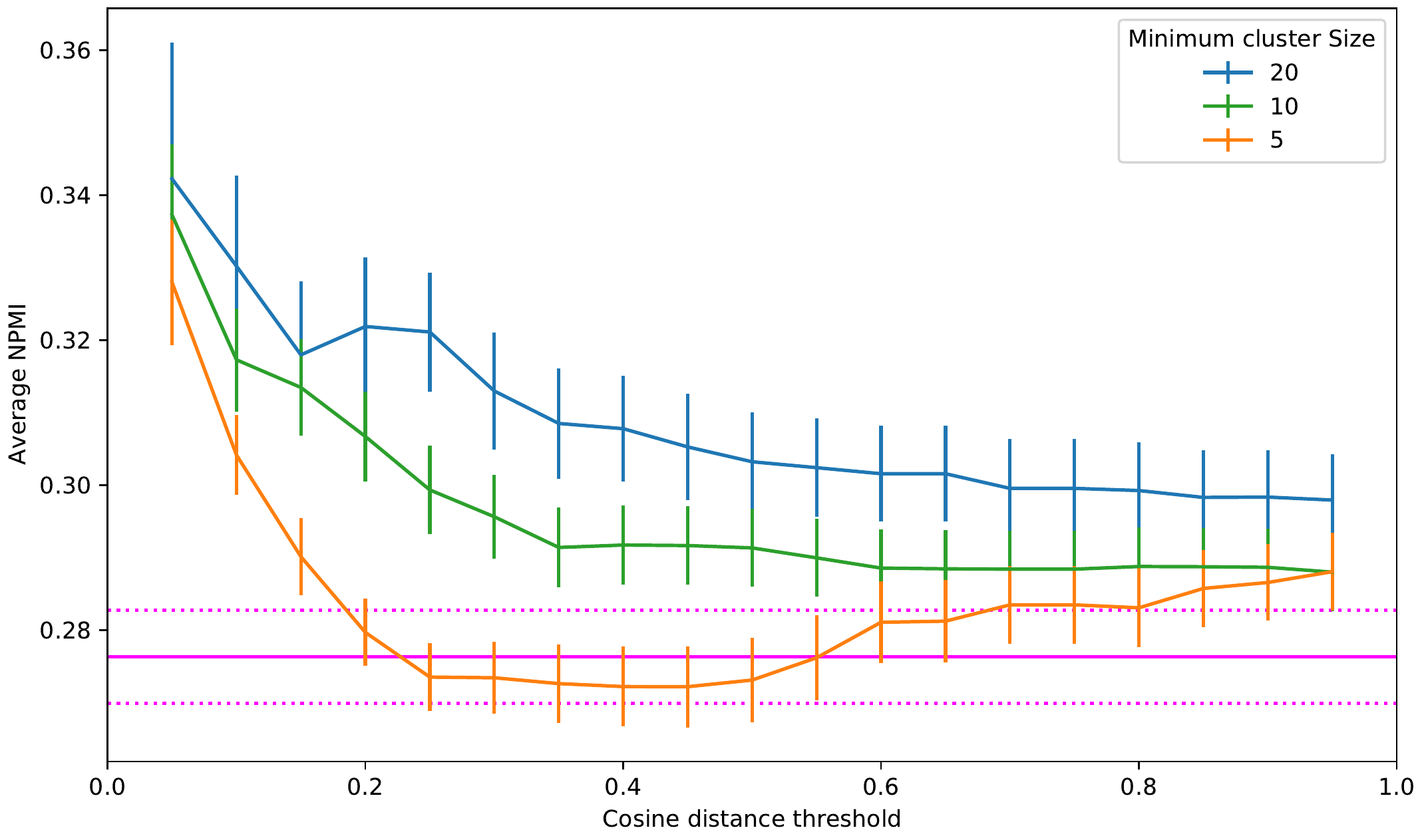}                           
            \label{Performance_2}
        \end{subfigure} 
        \\
  \begin{subfigure}[b]{0.48\textwidth}
            \centering 
            \caption{Distinctiveness}
            \includegraphics[width=1\linewidth]{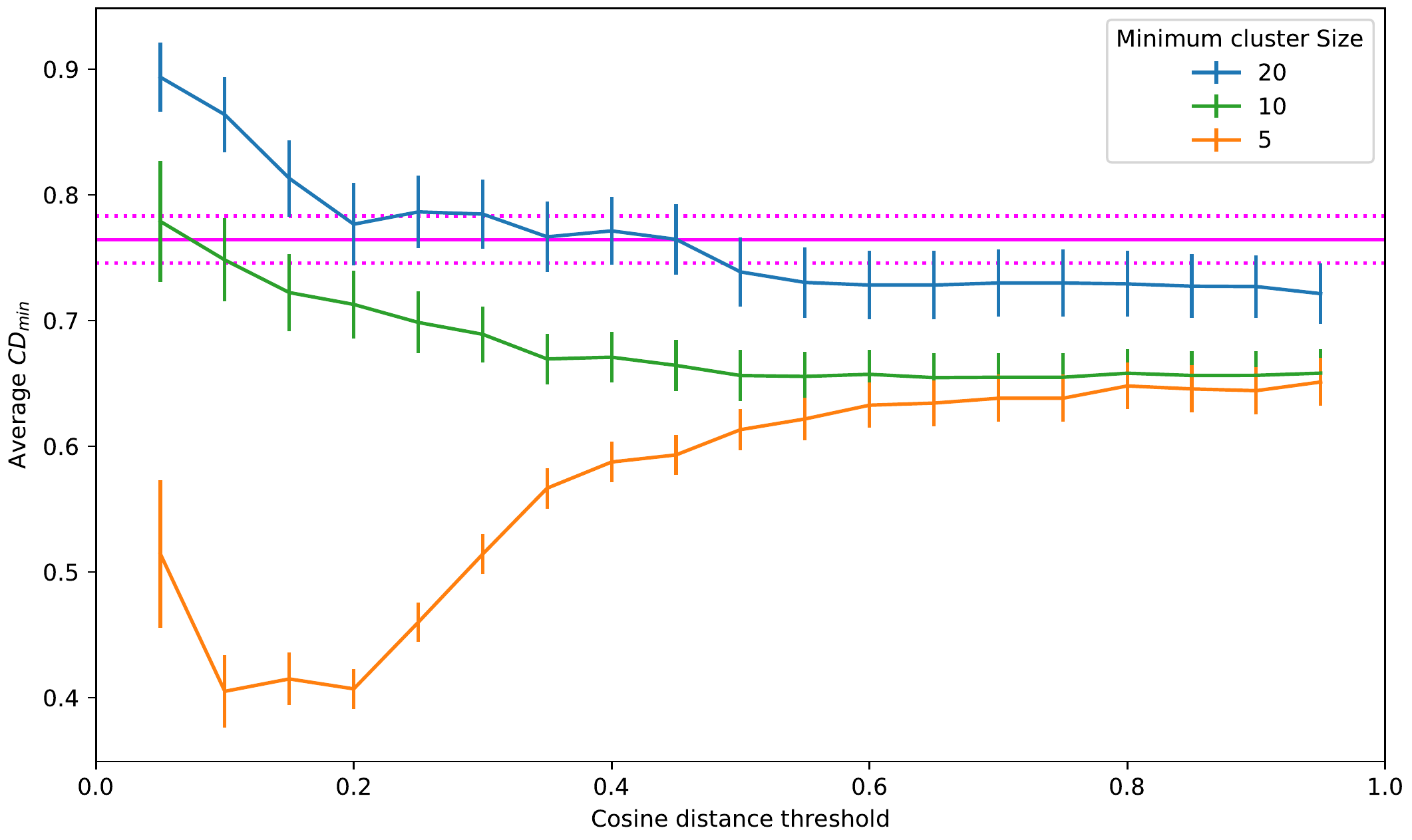}
             \label{Performance_3}           
        \end{subfigure}            
    \begin{subfigure}[b]{0.48\textwidth}
            \centering 
            \caption{Credibility}
            \includegraphics[width=1\linewidth]{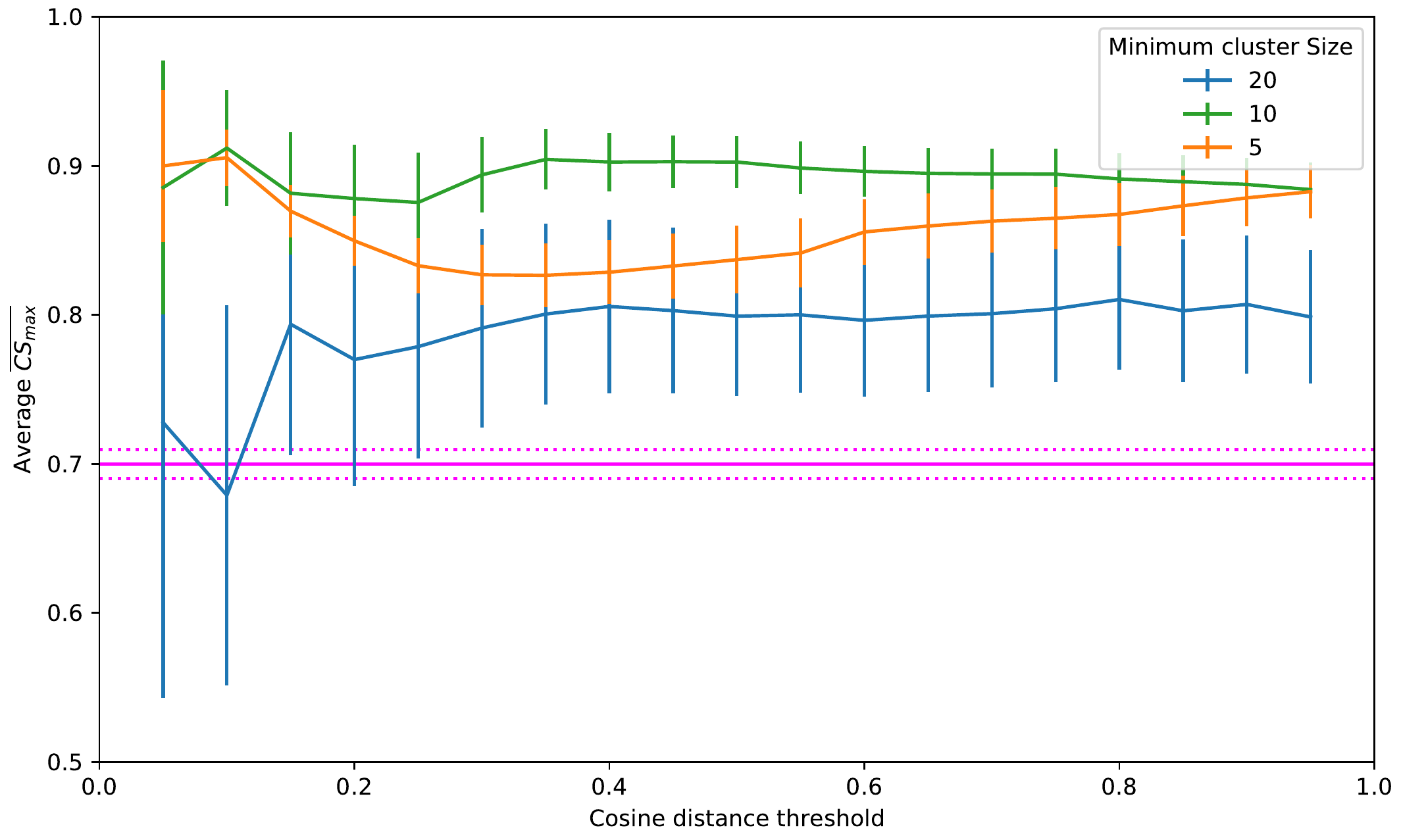}
             \label{Performance_4}           
        \end{subfigure}                     
        \caption{Evaluation of subsets of clustered topics. Subsets are formed with combinations of minimum cluster size and cosine distance thresholds. \textcolor{magenta}{Horizontal lines} and \textcolor{magenta}{dotted lines} show the average measures (± one standard error) of the STM posterior samples. Subsets of clusters formed with a minimum cluster size of 10 show greater coherence and credibility, and the subsets formed with a cosine distance threshold larger than 0.3 show better generalisation (in comparison to the average generalisation of the STM posterior samples). Subsets with a minimum cluster size of 10 show less distinctive clustered topics, which might result from filtering out distinctive but uncertain topics. A cosine distance threshold larger than 0.35 cosine distance does not significantly improve perplexity.}
        \label{Clustering}
\end{figure}

\section{Block Gibbs sampler}
\label{BGS}

\cite{chen2011sampling} proposes a block Gibbs sampling algorithm that jointly samples topic assignments and table indicators, leading to a more efficient sampling method. Table counts are not sampled, instead reconstructed by summation of the table indicators.

\begin{equation}
\label{eq:tablecounts}
\begin{aligned}
t_k = \sum^N_{n=1} u_n 1_{z_n=k},
\end{aligned}
\end{equation}

Using the table indicator representation, the PDP posterior distribution is:

\begin{equation}
\label{eq:PDP_ind}
\begin{aligned}
p(z,t \mid a,b,\theta) = \prod_k \frac{n_k!}{t!(n-t)!} p(z,u\mid a,b,\theta),
\end{aligned}
\end{equation}
responding to \(\frac{n_k!}{t!(n-t)!}\) sitting arrangements.

The joint distribution of topic assignments and table indicators can be obtained by
using Equation \ref{eq:PDP_ind} in Equation \ref{eq:STMlikelihood} resulting in:

\begin{equation}
\label{eq:STMjoint}
\begin{aligned}
&p(\mathbf{z},\mathbf{w},\mathbf{t} \mid \boldsymbol{\alpha},\boldsymbol{\beta},a,b) = \\
&\prod_d\frac{\textrm{Beta}_K(\boldsymbol{\alpha}+\sum_p\mathbf{t}_{p,d})}{\textrm{Beta}_K(\boldsymbol{\alpha})}\prod_{p,d}\frac{(b|a)_{\sum_k t_{p,d,k}}}{(b)_{N_{p,d}}}\prod_{p,d,k} S^{N_{k|p,d}}_{t_{p,d,k},a} \frac{t_{p,d,k}!(N_{p,d,k}-t_{p,d,k})!}{n_{p,d,k}!} \prod_k \frac{\textrm{Beta}_V(\boldsymbol{\beta}+\mathbf{N}_k)}{\textrm{Beta}_V(\boldsymbol{\beta})}
\end{aligned}
\end{equation}

The block Gibbs sampling algorithm firstly samples a table indicator \(u_{z_n} = 1\) or \(u_{z_n} = 0\) with probabilities:

\begin{equation}
\label{eq:indProb}
\begin{aligned}
p(u_{z_n} = 1 \mid z_n =k) = \frac{t_k}{n_k} \quad p(u_{z_n} = 0| z_n =k) = 1-\frac{t_k}{n_k},
\end{aligned}
\end{equation}
and discounts the current assignment \(z_n\) from \(N_{p,d,k}\) and reduces \(t_{p,d,k}\) by 1 if \(u_{z_n} = 1\).


Then, the full conditional distribution is computed taking into account two scenarios: the probability of opening a new table (Equation \ref{eq:blockgibbs1}) and the probability of choosing an occupied table (Equation \ref{eq:blockgibbs2}) if \(t_{p,d,k}^\prime > 0\).

\begin{equation}
\label{eq:blockgibbs1}
\begin{aligned}
&p(z_n=k, u_n=1 \mid \mathbf{z}-\lbrace z_{n} \rbrace, \mathbf{u}-\lbrace u_{n}\rbrace ,\mathbf{w},\boldsymbol{\alpha},\boldsymbol{\beta},a,b) \propto \\
&\frac{\alpha_k+t_{d,k}^\prime}{\alpha+t_d^\prime}
\frac{b+a  t_{p,d}^\prime}{b+N_{p,d}^\prime} 
\frac{S^{N_{p,d,k}^\prime+1}_{t_{p,d,k}^\prime+1}}{S^{N_{p,d,k}^\prime}_{t_{p,d,k}^\prime}} \frac{t^\prime_{p,d,k}+ 1.0}{n^\prime_{p,d,k} + 1.0}\frac{\beta_v+M_{k,w_{p,d,n}}^\prime}{\beta+M_k^\prime},
\end{aligned}
\end{equation}

\begin{equation}
\label{eq:blockgibbs2}
\begin{aligned}
&p(z_n=k, u_n=0|\mathbf{z}-\lbrace z_{n} \rbrace, \mathbf{u}-\lbrace u_{n}\rbrace ,\mathbf{w},\boldsymbol{\alpha},\boldsymbol{\beta},a,b) &\propto \\
&\frac{S^{N_{p,d,k}^\prime+1}_{t_{p,d,k}^\prime}}{S^{N_{p,d,k}^\prime}_{t_{p,d,k}^\prime}} \frac{1}{b+N^\prime_{p,d}}\frac{n_{p,d,k}^\prime-t _{p,d,k}^\prime+ 1.0}{n_{p,d,k}^\prime + 1.0} \frac{\beta_v+M_{k,w_{p,d,n}}^\prime}{\beta+M_k^\prime},
\end{aligned}
\end{equation}
where the dash indicates statistics after excluding the current assignment.

Finally, update the counts of \(n_{p,d,k}\) and \(t_{p,d,k}\) with the sampled topic assignment \(z_n\) and table indicator \(u_n\).

\section{Hierarchical clustering}\label{HC}

The hierarchical clustering algorithm takes a bag of topics, a list with sample indexes, and a cosine distance threshold. The bag of topics gathers topic distributions from various posterior samples from various MCMC. The list of sample indices records a sample index for each topic, i.e., assuming that the first 50 topics in the bag of topics come from posterior sample 1 and the next 50 topics come from posterior sample 2, then the first 50 elements in the list of samples indices are 1 and the next 50 elements are 2. The cosine distance threshold indicates the limit up to which topics would be merged. 

The algorithm will start by forming clusters with each of the topics in the bag of topics. So, if there are \(N\) topics, there are \(N\) initial clusters. Then, a list \(L\) is created to record the cosine distance between two clusters. This list contains the indexes of the two compared clusters and the cosine distance between the clustered topics. A clustered topic is the average topic distributions of the cluster members.

At each step, the algorithm finds the pair of clusters in \(L\) with the minimum cosine distance. Then, the algorithm evaluates if the members of both clusters are from different posterior samples using the list of sample indices. If so, a new cluster is created by merging the evaluated pair of clusters. Then, the algorithm removes from the \(L\) all comparisons that had any of the identified clusters and adds comparisons from all the remaining clusters to the new cluster. But, If the evaluation is false, the algorithm updates the cosine distance between the pair of clusters with 1. Thereby, the algorithm would not take the same pair of clusters in the next step.

The algorithm will keep merging clusters until the minimum cosine distance is larger than the cosine distance threshold. The algorithm then retrieves all the remaining clusters (clusters that are not eliminated because they do not get merged).

\end{appendix}
    
\bibliographystyle{imsart-nameyear} 
\bibliography{references}

\end{document}